\documentclass[twocolumn,twocolappendix,resetfootnote]{aastex701}

\usepackage{amsmath}

\newcommand{\ELISA}{\texttt{ELISA}}
\newcommand{\XSPEC}{\texttt{XSPEC}}

\begin{document}

\title{\texttt{ELISA}: A Bayesian X/$\gamma$-ray Spectral Analysis Toolkit}

\newcommand{\SKPLA}{State Key Laboratory of Particle Astrophysics, Institute of High Energy Physics, Chinese Academy of Sciences, Beijing 100049, China}
\newcommand{\UCAS}{University of Chinese Academy of Sciences, Chinese Academy of Sciences, Beijing 100049, China}

\correspondingauthor{Sheng-Lun Xie, Shao-Lin Xiong, Xiao-Bo Li}
\email{xiesl@jgsu.edu.cn, xiongsl@ihep.ac.cn, lixb@ihep.ac.cn}

\author[0000-0001-8664-5085]{Wang-Chen Xue}
\affiliation{\SKPLA}
\email{xuewc@ihep.ac.cn}

\author[0000-0001-9217-7070]{Sheng-Lun Xie\textsuperscript{*}}
\affiliation{Institute for Astronomy and Astrophysics, Jinggangshan University, Ji'an, Jiangxi 343009, China}
\affiliation{Key Laboratory of Energy Conversion Optoelectronic Functional Materials of Jiangxi Education Institutes, School of Mathematics and Physics, Jinggangshan University, Ji'an, Jiangxi 343009, China}
\email{xiesl@jgsu.edu.cn}

\author[0009-0001-7226-2355]{Chao Zheng}
\affiliation{\SKPLA}
\affiliation{TIANFU Cosmic Ray Research Center, Chengdu, Sichuan, China}
\email{zhengchao_astro@foxmail.com}

\author[0000-0002-4771-7653]{Shao-Lin Xiong\textsuperscript{*}}
\affiliation{\SKPLA}
\email{xiongsl@ihep.ac.cn}

\author[0000-0003-4585-589X]{Xiao-Bo Li\textsuperscript{*}}
\affiliation{\SKPLA}
\email{lixb@ihep.ac.cn}

\author[0000-0001-9834-2196]{Yong Chen}
\affiliation{\SKPLA}
\email{ychen@ihep.ac.cn}

\author[0000-0002-8097-3616]{Peng Zhang}
\affiliation{\SKPLA}
\affiliation{College of Electronic and Information Engineering, Tongji University, Shanghai 201804, China}
\email{zhangp97@ihep.ac.cn}

\author[0000-0001-5348-7033]{Yan-Qiu Zhang}
\affiliation{\SKPLA}
\affiliation{\UCAS}
\affiliation{School of Physics and Electronic Science, Guizhou Normal University, Guiyang 550001, China}
\email{yanqiuzhang@gznu.edu.cn}

\author[0009-0004-1887-4686]{Jia-Cong Liu}
\affiliation{\SKPLA}
\affiliation{\UCAS}
\email{liujiacong@ihep.ac.cn}

\author[0009-0008-8053-2985]{Chen-Wei Wang}
\affiliation{\SKPLA}
\affiliation{\UCAS}
\email{cwwang@ihep.ac.cn}

\author[0009-0006-5506-5970]{Wen-Jun Tan}
\affiliation{\SKPLA}
\affiliation{\UCAS}
\email{tanwj@ihep.ac.cn}

\author[0009-0008-5068-3504]{Yue Wang}
\affiliation{\SKPLA}
\affiliation{\UCAS}
\email{yuewang@ihep.ac.cn}

\author[0009-0007-6192-0213]{Jin-Peng Zhang}
\affiliation{\SKPLA}
\affiliation{\UCAS}
\email{zhangjinpeng@ihep.ac.cn}

\author[0009-0002-6411-8422]{Zheng-Hang Yu}
\affiliation{\SKPLA}
\affiliation{\UCAS}
\email{zhyu@ihep.ac.cn}

\begin{abstract}
Forward-folded X/$\gamma$-ray spectral fitting increasingly benefits from Bayesian workflows to better capture counting statistics, background uncertainties, and complex parameter degeneracies,
while fully reproducible end-to-end implementations are still not always straightforward to deploy in routine analyses.
We present \ELISA\ (Efficient Library for Spectral Analysis), an open-source Python toolkit that streamlines the full spectral analysis loop for X/$\gamma$-ray astronomy: data loading, model construction, inference, and model validation and comparison. \ELISA\ couples a \texttt{JAX}-accelerated forward model with a unified fitting interface, supporting both rapid optimization and modern Bayesian sampling.
Built-in diagnostics make convergence checks and predictive assessment a standard part of the workflow, motivating principled model refinement when needed.
Using a spectral fitting example of the \textit{Insight}-HXMT Crab observation, we find substantial improvements in sampling efficiency compared with \XSPEC's default MCMC implementation. \ELISA\ is available at \url{https://github.com/wcxve/elisa} and on PyPI at \url{https://pypi.org/project/astro-elisa}.
\end{abstract}


\section{Introduction}\label{sec:intro}
X/$\gamma$-ray spectroscopy is a primary means of probing the physical conditions of high-energy sources, including accretion flows, relativistic outflows, and transient emission from compact objects.

For most high-energy instruments, however, the measured data are not a direct sampling of the incident photon spectrum, but discrete photon counts accumulated in detector channels after redistribution by the instrumental response and other observation-specific effects. Recovering the incident spectrum by directly inverting this response is generally an ill-posed and numerically unstable problem, and does not yield a unique, model-independent solution.
Scientific inference, therefore, relies on \emph{forward folding}: an astrophysical spectral model is mapped to expected counts by convolution with the instrumental response and any other observation-specific effects, and the model parameters are constrained by comparing predicted and observed counts through an appropriate counting likelihood.
This process tightly couples astrophysical modeling, calibration products, and statistical inference, which makes practical analysis sensitive to both numerical performance and methodological robustness.

In practice, forward-folded spectral analysis is typically carried out with a small set of mature toolkits. \XSPEC\ \citep{Arnaud1996} remains the de facto standard, offering an extensive, well-validated model library, broad mission support, and a robust suite of established fit statistics.
\texttt{Sherpa} \citep{Freeman2001Sherpa, Doe2007Sherpa, Siemiginowska2024Sherpa} offers a Python-native environment with flexible model composition and optimization.
\texttt{SPEX} \citep{Kaastra2024SPEX} is widely used for fitting high-resolution X-ray spectroscopy.
In addition, the Interactive Spectral Interpretation System (\texttt{ISIS}) \citep{2000ASPC..216..591H} also provides a flexible environment for high-resolution X-ray spectral analysis, with strong support for spectral-line identification, plasma-emission modeling, and access to atomic databases such as APED.
\texttt{Gammapy} \citep{2023A&A...678A.157D} provides an open-source Python framework for astronomical $\gamma$-ray data analysis, supporting event and instrument-response-based data reduction, background estimation, Poisson likelihood fitting, and spectral, temporal, and spatial modeling, including multi-instrument analyses.
These tools have enabled high-impact results and provide important foundations for modern high-energy spectral analysis. Here, we present a complementary implementation aimed at integrating forward-folded spectral modeling, Bayesian posterior inference, posterior predictive checking, and model comparison within a single reproducible workflow.

The demand for such workflows is increasingly driven by the scale and complexity of modern high-energy astrophysical data, as well as by the more stringent questions they are now used to address.
Bayesian approaches have long been recognized as valuable in high-energy astrophysics, for example, in the treatment of low-count X-ray spectral analysis and related inference problems \citep[e.g.,][]{2001ApJ...548..224V, 2011ASPC..442..439S}.
Many analyses now operate in regimes where (i) the Poisson nature of counts and the treatment of background dominate the uncertainty budget; (ii) parameter posteriors are non-Gaussian, multimodal, or strongly correlated; and (iii) scientific conclusions depend on model adequacy and predictive performance.
Bayesian approaches address these issues naturally, but in daily practice, they remain challenging because interfaces across modeling, sampling, and diagnostics are often fragmented, and the computational cost of repeated forward-folding can be prohibitive without a modern acceleration technique.

There are a number of recent efforts that have substantially improved the landscape. Within the \XSPEC\ ecosystem, \texttt{BXA} \citep{Buchner2014BXA, Buchner2021BXA} provides a practical bridge from \XSPEC\ models to Bayesian inference by interfacing nested sampling engines (UltraNest; \citealt{Buchner2016UltraNest, Buchner2019UltraNest, Buchner2021UltraNest}), enabling posterior exploration and evidence-based comparison. More broadly, \texttt{3ML} \citep{Vianello2015ThreeML} supplies a modular likelihood layer that facilitates joint inference across missions and data types, supporting both frequentist and Bayesian analyses.
In parallel, spectral modeling projects (e.g., \texttt{jaxspec}; \citealt{Dupourque2024jaxspec}) demonstrate that just-in-time compilation and automatic differentiation based on \texttt{JAX} \citep{Bradbury2018JAX} can accelerate repeated model evaluations and open the door to gradient-informed inference in spectral problems.
Together, these developments highlight both the demand for Bayesian workflows and the feasibility of modern high-performance implementations.

Despite methodological and software progress, routine Bayesian spectral analysis still lacks a standardized workflow in which model building, fitting, validation, and comparison are treated as equally essential stages \citep{Gelman2020Workflow}.
In practice, analysts are often forced to balance the maturity of mission-hardened spectral ecosystems (e.g., \XSPEC) against adopting Bayesian tooling that more naturally supports systematic diagnostics and predictive evaluation.
Even when posterior sampling is available, post-fit steps are frequently assembled in an ad hoc fashion.
Specifically, posterior predictive checks \citep{Gelman1996PPC} and simulation-calibrated goodness-of-fit procedures are seldom integrated into the main analysis loop, which is critical in common non-regular settings where reference distributions can fail \citep{protassov_statistics_2002}.
Likewise, principled predictive model evaluation via leave-one-out cross-validation (e.g., PSIS-LOO; \citealt{Vehtari2015LOO}) and information criteria (e.g., WAIC; \citealt{Watanabe2010WAIC}) is often performed outside the fitting pipeline, rather than as a routine, reproducible component of model comparison.

To address these needs, we developed \ELISA\ (Efficient Library for Spectral Analysis; \citealt{ELISA}), an open-source Python toolkit that makes statistically rigorous, end-to-end Bayesian spectral analysis practical for routine work.
Since its public release, \ELISA\ has been used in spectral analyses of data from several high-energy missions, including GECAM \citep{GECAM1, GECAM2}, \textit{Insight}-HXMT \citep{Zhang20HXMTOverview}, SVOM/GRM \citep{SVOM_1, SVM_GRM_Dong2010}, and \textit{Fermi}/GBM \citep{GBM_meegan_09}, and applied to a broad range of high-energy sources in the literature (e.g., \citealt{2025ApJ...983..130L,2025ApJ...987L...7L,2025SCPMA..6871011Z,2025arXiv250926283Z,2026ApJ...999..159X,2026ApJ..1003..136X,2025ApJ...993...24W,2025arXiv251215162L,2025ApJ...987L..38Z}). It has also been evaluated and incorporated into internal and collaborative analysis pipelines (e.g., \citealt{2025arXiv251015816W}). These applications have shaped \ELISA's design priorities, including robust defaults, transparent diagnostics, and performance computing backends.

While \ELISA\ supports both frequentist and Bayesian inference, this paper focuses on the Bayesian workflow.
The remainder of this paper is organized as follows.
Section~\ref{sec:overview} outlines the end-to-end workflow of \ELISA\ and the design goals.
Section~\ref{sec:data_and_model} describes OGIP data handling and model construction interfaces.
Section~\ref{sec:stats} formulates the statistical model for forward-folded fitting, including likelihood choices and background treatment.
Section~\ref{sec:inference} introduces the posterior inference methods.
Section~\ref{sec:diagnostics} describes the tools for model checking and comparison.
We conclude in Section~\ref{sec:conclusion} with current limitations and future directions.

\section{Overview and Design Goals}\label{sec:overview}

\ELISA\ is designed to support forward-folded X/$\gamma$-ray spectral analysis within a modern Bayesian workflow, spanning model specification and inference as well as principled model checking and comparison.
This section summarizes the intended analysis loop and the core design goals.

\subsection{End-to-end workflow}\label{sec:overview:workflow}

\begin{figure}
\centering
\includegraphics[width=\columnwidth]{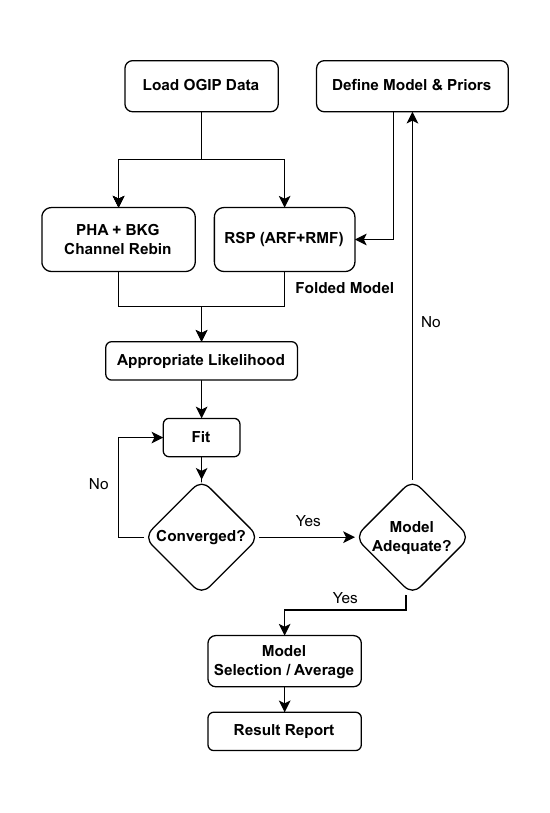}
\caption{Schematic end-to-end Bayesian workflow implemented by \ELISA.}
\label{fig:workflow}
\end{figure}

Figure~\ref{fig:workflow} sketches the analysis loop targeted by \ELISA.
For a typical analysis, one first loads one or more datasets and performs data preparation, including channel regrouping, energy filtering, and bad-channel masking. In particular, users can define the energy range used in the fit, while channels flagged through the OGIP \texttt{QUALITY} column are consistently excluded.
One then defines a spectral model, then chooses a likelihood appropriate for binned counting data and the adopted background treatment, then performs inference, either rapid maximum-likelihood estimation for prototyping, or full Bayesian posterior inference for uncertainty quantification, and finally evaluates model adequacy and performance through standardized diagnostics and model comparison procedures.
These diagnostics can motivate revisions to the model structure, priors, or likelihood assumptions, closing the iterative loop shown in Figure~\ref{fig:workflow}.

\subsection{Design goals}\label{sec:overview:goals}

\ELISA\ is designed around the following goals:
\begin{itemize}
\item Workflow-first spectral analysis: treat model building, inference,
diagnostics, and comparison as a single, reproducible workflow rather than disconnected steps.
\item Operational reproducibility: by ``reproducible'' we mean that the same analysis can be rerun from versioned inputs, explicit energy selections and grouping choices, declared model/prior definitions, and recorded inference settings to recover the same scientific products up to expected numerical tolerances.
\item Composable models: build models by combining modular components, supporting expressive constructions and parameter linking across components and datasets.
\item Unified inference: support both rapid maximum-likelihood fits and full Bayesian posterior inference within a consistent framework.
\item Standardized model checking and comparison: provide built-in tools for model checking and model comparison, emphasizing calibrated assessment of model adequacy and performance.
\item \texttt{JAX} backend for performance and differentiability: implement the forward model and likelihood in \texttt{JAX} \citep{Bradbury2018JAX} to enable just-in-time compilation and automatic differentiation.
\end{itemize}

\section{Spectral Data and Model Interface}\label{sec:data_and_model}

\subsection{Data Loading Interface}\label{sec:data_and_model:data}

\ELISA\ is designed to work compatibly with OGIP-standard\footnote{
    See \textit{``The OGIP Spectral File Format''} at \url{https://heasarc.gsfc.nasa.gov/docs/heasarc/ofwg/docs/spectra/ogip_92_007/ogip_92_007.html}, and \textit{``The Calibration Requirements for Spectral Analysis (Definition of RMF and ARF file formats)''} at \url{https://heasarc.gsfc.nasa.gov/docs/heasarc/caldb/docs/memos/cal_gen_92_002/cal_gen_92_002.html}.
} spectral products produced by mission pipelines and existing tools: a source spectrum (PHA), calibration products (RSP, or RMF with an optional ARF), and an optional background spectrum (PHA).
Both type~I and type~II PHA formats are supported, including multi-spectrum products.
\ELISA\ also accepts OGIP spectra stored either as counts (\texttt{COUNTS}) or as count rates (\texttt{RATE}/\texttt{RATES}); rate-form inputs are converted using the recorded exposure, after which only likelihood choices consistent with the implied counting or Gaussian error model are appropriate.
Associated files can be identified through OGIP header keywords (\texttt{RESPFILE}, \texttt{ANCRFILE}, \texttt{BACKFILE}) when present.
Energy selections are mapped to detector channels using the response \texttt{EBOUNDS} table, and channels flagged by bad \texttt{QUALITY} are excluded consistently.

\ELISA\ performs joint grouping of the spectrum, background, and response so that forward-folded model evaluations are defined on the fitted channel grid.
Its grouping behavior is intermediate between the usual \XSPEC\ and \texttt{Sherpa} workflows: when OGIP \texttt{GROUPING} and \texttt{QUALITY} information is present in the input files, \ELISA\ honors these file-level definitions, while still allowing the user to modify the selected channels and regroup the data after loading. Post-loading regrouping is applied consistently to the source spectrum, background spectrum, and response.
\ELISA\ provides several adaptive rebinning schemes commonly used in high-energy spectroscopy, including fixed-width binning in channel space, minimum total counts per bin, and minimum source significance.
\ELISA\ also implements response-aware optimal binning \citep{Kaastra2016OptBin}, using the instrumental resolution to avoid overbinning at high resolution while maintaining adequate counts where needed.
For Poisson spectra with background data provided, \ELISA\ provides background-aware constraints motivated by methods that treat the background explicitly by profiling over the background likelihood (e.g., W-stat and PG-stat; \citealt{Wachter1979Wstat,Arnaud1996}). In this setting, the background spectrum is required, after rebinning, to meet minimum per-channel count or relative-uncertainty criteria to mitigate statistical bias in the inferred source properties (see Section~\ref{sec:stats:wstat} and Section~\ref{sec:stats:pgstat}). The interface for loading OGIP data is described in more detail in \ELISA's online API documents\footnote{\url{https://astro-elisa.rtfd.io/en/stable/apidoc/elisa.data.ogip.html\#elisa.data.ogip.Data}}.

\subsection{Model Construction Interface}\label{sec:data_and_model:model}

\ELISA\ represents spectral models in photon space as compositional expressions built from additive emission components, multiplicative modifiers, and convolution operators.
Each component defines a photon spectrum and provides an evaluation routine over an energy grid; for photon flux computed over a given energy range, analytic integrals are used when available, while numerical integration (e.g., trapezoidal or Simpson rules) is applied otherwise.
Convolution operators wrap submodels to implement transformations such as redshift or flux normalization while preserving a unified model representation.
\ELISA\ provides native implementations of several commonly used spectral components, including continuum models such as power laws, cutoff power laws, Band-like functions, and blackbodies, together with simple additive features such as Gaussian lines. These components can be combined within the \ELISA\ model interface to build composite spectral models. The model interface is also extensible: users can define various custom spectral components directly in Python, and models written with \texttt{JAX}-compatible operations can take advantage of just-in-time compilation and automatic differentiation.
In addition, \ELISA\ can interoperate with the \XSPEC\ extensive model library through our \texttt{Python} wrapper package \texttt{xspex}\footnote{\url{https://github.com/wcxve/xspex}}, which exposes \XSPEC\ models to Python and integrates them into the \ELISA\ model interface. For external components that are not natively written in \texttt{JAX}, such as \XSPEC\ models, automatic differentiation is enabled by finite-difference Jacobian–vector products, allowing them to be used within the \ELISA\ workflow.

Model parameters are represented as objects that store a default value used for initialization, and a supported range used both as an optimization constraint and as the default prior support for Bayesian inference.
By default, \ELISA\ assigns bounded uniform priors (or log-uniform priors when a
parameter is declared on a logarithmic scale), while allowing parameters to be
fixed to constants or assigned alternative priors via arbitrary distributions from \texttt{NumPyro} \citep{phan2019composable, bingham2019pyro}.
Parameter objects can be shared across components and datasets to implement tying, and deterministic parameter relations can be expressed by composing parameters algebraically.

For more details on the model building interface, we refer the interested user to \ELISA's online tutorial\footnote{\url{https://astro-elisa.rtfd.io/en/stable/notebooks/model-building.html}}.

\section{Statistical Framework for Forward-Folded Model}\label{sec:stats}

This section summarizes the statistical framework implemented in \ELISA.
The implementation is designed around a forward-folded generative model for
binned spectra, with likelihood options chosen to match the data type and the
available background information.
In the current implementation, background contributions are handled either as fixed expectations, profiled nuisance terms, or Gaussian-treated net/background measurements; \ELISA\ does not yet support attaching a separate forward-folded physical background model with its own response as part of the same observation object.

\subsection{Notation and forward-folded source model}\label{sec:stats:forward}

We consider one or more datasets indexed by $d=1,\ldots,D$.
Conditioned on the model parameters (including any shared nuisance parameters),
the datasets are assumed statistically independent.
Accordingly, the present formulation does not directly represent cases with explicit cross-dataset covariance structures, such as joint spectral likelihoods for correlated Stokes products (e.g., IXPE-style $I/Q/U$ analyses).
After energy selection and grouping in channel space, dataset $d$ contains
$J_d$ channels indexed by $j$, with observed counts in the source (``on'')
spectrum denoted by $N_{dj}$.
Here, the index $j$ refers to the fitted grouped channels (or bins) after all channel masking and regrouping steps, not necessarily to the native PI/PHA channels in the original file.
If a background (``off'') spectrum is available, the background counts are
$N^{\mathrm{off}}_{dj}$.
We denote by $\alpha_d$ the (dataset-level) scaling factor that maps the
off-background expectation to the on-background expectation, i.e.,
$\mathbb{E}[B_{dj}] = \alpha_d\,\mathbb{E}[B^{\mathrm{off}}_{dj}]$.
For OGIP products, $\alpha_d$ is constructed from the exposure and the
\texttt{AREASCAL}/\texttt{BACKSCAL} metadata.
More generally, this scaling may be channel dependent, in which case one may write $\alpha_{dj}$; this arises, for example, when OGIP \texttt{AREASCAL} or \texttt{BACKSCAL} are stored as vectors rather than scalars. For notational simplicity, we retain $\alpha_d$ in the equations below when the channel dependence is either absent or understood implicitly.

Let $M_d(E\mid\theta_d)$ be the dataset-specific photon-space spectral model
(e.g., in units of ph\,cm$^{-2}$\,s$^{-1}$\,keV$^{-1}$), where $\theta_d$
represents the free parameters for dataset $d$.
Models may be shared across datasets, and parameters can be linked or constrained
through deterministic relations, depending on the analysis goals.
For each dataset, the model is evaluated on the photon energy grid with edges
$E_{di}$ (as recorded by the response), and the integrated model in energy bin
$i$ is
\begin{equation}
  m_{di}(\theta_d) \equiv \int_{E_{di}}^{E_{d,i+1}} M_d(E \mid \theta_d)\, dE .
\end{equation}
The instrument response is represented by an operator $R_{dji}$ constructed
from the OGIP RMF and (optionally) ARF products, such that the expected source
count rate in channel $j$ is
\begin{equation}
  r_{dj}(\theta_d) = \sum_i R_{dji}\, m_{di}(\theta_d),
\end{equation}
and the expected source counts are
\begin{equation}
  s_{dj}(\theta_d) = t_d\, r_{dj}(\theta_d),
\end{equation}
where $t_d$ is the exposure time of the on spectrum.
In matrix form, $\mathbf{s}_d(\theta_d)=t_d\,\mathbf{R}_d\,\mathbf{m}_d(\theta_d)$.

\subsection{Likelihood choices}\label{sec:stats:likelihood}

Let $\Theta$ denote the global parameter vector.
Dataset-specific parameters $\theta_d$ are obtained via deterministic relations
(including parameter linking), i.e., $\theta_d=g_d(\Theta)$, and we write
$s_{dj}(\Theta)$ for brevity.
\ELISA\ assumes channel-wise conditional independence given the expected counts
(and, when applicable, background terms), so the likelihood factorizes over $j$.
For numerical stability and for consistency with common fit-statistic practice (as in \XSPEC), the implemented ``statistics'' correspond to log likelihoods up to additive constants that depend only on the observed data.
Table~\ref{tab:likelihood-options} summarizes the likelihood options for common spectral data configurations.
By default, \ELISA\ automatically selects an appropriate likelihood for each dataset based on the data type and the available background information, while still allowing users to override this choice when required by the analysis.

\begin{deluxetable}{ccc}

    \setlength{\tabcolsep}{2pt}
    \tablewidth{0pt}
    \tablecaption{Likelihood options for common spectral data configurations.\label{tab:likelihood-options}}
    \tablehead{
      \colhead{Data type} &
      \colhead{Background type} &
      \colhead{Likelihood (Statistic)}
    }
    \startdata
    Poisson & None & Poisson (C-stat) \\
    Poisson & Known expectation & Poisson (P-stat) \\
    Poisson & Poisson & Poisson-Poisson (W-stat) \\
    Poisson & Gaussian & Poisson-Gaussian (PG-stat) \\
    Gaussian & None, or Gaussian & Gaussian ($\chi^2$) \\
    \enddata
\end{deluxetable}

\subsubsection{Poisson likelihood without background (C-stat)}\label{sec:stats:cstat}

For Poisson-count spectra without an explicit background term, the generative
model is
\begin{equation}
  N_{dj} \sim \mathrm{Poisson}\!\left(s_{dj}(\Theta)\right),
\end{equation}
with log likelihood (dropping $\ln N_{dj}!$)
\begin{equation}
  \ln \mathcal{L}_d(\Theta) =
  \sum_{j=1}^{J_d} \left[
    N_{dj}\ln s_{dj}(\Theta) - s_{dj}(\Theta)
  \right].
\end{equation}
This corresponds to the Cash statistic \citep{Cash1979Cstat} up to an additive constant.

\subsubsection{Poisson likelihood with known background (P-stat)}\label{sec:stats:pstat}

When a background expectation is treated as known, \ELISA\ uses
\begin{equation}
  N_{dj} \sim \mathrm{Poisson}\!\left(s_{dj}(\Theta) + \alpha_d \mu^{\mathrm{off}}_{dj}\right),
\end{equation}
where $\mu^{\mathrm{off}}_{dj}$ denotes the known off-background expectation in
channel $j$.
In the high-count regime, one may approximate
$\mu^{\mathrm{off}}_{dj}\approx N^{\mathrm{off}}_{dj}$ as a plug-in estimate,
which is appropriate when the uncertainty in the background estimate is
negligible compared to the source uncertainty, or when the background has been
constrained elsewhere and is used as a fixed input.
In \ELISA, this fixed-background treatment corresponds to an explicitly chosen likelihood option; it is not selected automatically by the default likelihood-dispatch logic.

\subsubsection{Poisson background with profiling (W-stat)}\label{sec:stats:wstat}

For Poisson on/off spectra, \ELISA\ supports a profile-likelihood treatment
commonly known as W-stat \citep{Wachter1979Wstat, Arnaud1996}.
Per-channel nuisance parameters $b_{dj} \ge 0$ are introduced to represent the expected off-background counts, and the likelihood is formulated as
\begin{equation}
\begin{aligned}
  N_{dj} &\sim \mathrm{Poisson}\!\left(s_{dj}(\Theta) + \alpha_d b_{dj}\right), \\
  N^{\mathrm{off}}_{dj} &\sim \mathrm{Poisson}\!\left(b_{dj}\right).
\end{aligned}
\end{equation}
Rather than sampling $\{b_{dj}\}$, the implementation profiles them out by
maximizing the joint likelihood channel-by-channel for each $\Theta$, enforcing
the non-negativity constraint $b_{dj}\ge 0$ and applying stable handling of
edge cases (e.g., zero on/off counts).
The closed-form profile solution $\hat{b}_{dj}(s_{dj})$ is given in
Appendix~\ref{app:profile-background:wstat}.
This yields an effective likelihood for $\Theta$ that retains the finite-count
uncertainty of both the on and off measurements.
In practice, in low-background-count regimes, a commonly used rule of thumb for
W-stat is to group channels such that the background spectrum has at least
$\sim 5$ counts per channel (i.e., $N^{\mathrm{off}}\gtrsim 5$) after
grouping, which improves the stability of the profile-likelihood background
handling and reduces the risk of biased source-parameter inference (e.g.,
\citealt{Buchner23SpecStat}).

\subsubsection{Gaussian background with profiling (PG-stat)}\label{sec:stats:pgstat}

When a background estimate is supplied with Gaussian uncertainties
$\sigma^{\mathrm{off}}_{dj}$ (e.g., for spectra treated as non-Poisson), \ELISA\
implements a profile-likelihood analogous to the PG-stat option in \XSPEC\ \citep{Arnaud1996}.
The on spectrum remains Poisson, while the background measurement is modeled as
Gaussian:
\begin{equation}
\begin{aligned}
  N_{dj} &\sim \mathrm{Poisson}\!\left(s_{dj}(\Theta) + \alpha_d b_{dj}\right), \\
  N^{\mathrm{off}}_{dj} &\sim \mathcal{N}\!\left(b_{dj},\,\sigma^{\mathrm{off}}_{dj}\right).
\end{aligned}
\end{equation}
For each $\Theta$, $b_{dj}$ is profiled out in closed form under the constraint
$b_{dj}\ge 0$, see Appendix~\ref{app:profile-background:pgstat}.
Similar to the W-stat treatment, when the background uncertainties $\sigma^{\mathrm{off}}_{dj}$ are large, we recommend grouping channels to reduce per-channel fractional background errors, e.g., requiring a minimum signal-to-uncertainty ratio in the background estimate such that $N^{\mathrm{off}}/\sigma^{\mathrm{off}}\gtrsim 2.5$ after grouping.
This typically strengthens the effective constraints on $b_{dj}$ in the profile-likelihood treatment and helps mitigate the risk of biased source-parameter inference.

\subsubsection{Gaussian likelihood (Chi-square)}\label{sec:stats:chi2}

For spectra treated with Gaussian uncertainties, \ELISA\ works with a net
spectrum and per-channel errors.
If a background spectrum is present, the net counts and their uncertainties are
\begin{equation}
  N^{\text{net}}_{dj} = N_{dj} - \alpha_d N^{\mathrm{off}}_{dj},
  \qquad
  \sigma^2_{dj} = \sigma^2_{N_{dj}} + \alpha_d^2 (\sigma^{\mathrm{off}}_{dj})^2,
\end{equation}
where $\sigma_{N_{dj}}$ and $\sigma^{\mathrm{off}}_{dj}$ are the per-channel
statistical errors provided by the OGIP PHA products (e.g., \texttt{STAT\_ERR}
when available; otherwise using the usual Poisson approximation when applicable),
optionally inflated by fractional systematics (the \texttt{SYS\_ERR} column in the PHA) for Gaussian-statistics fits.
The likelihood is then
\begin{equation}
  N^{\text{net}}_{dj} \sim \mathcal{N}\!\left(s_{dj}(\Theta),\,\sigma_{dj}\right).
\end{equation}
This option is intended for regimes where Gaussian approximations are adequate
(e.g., sufficiently large counts after grouping). We note, however, that
Gaussian-based fitting of Poisson data can still yield biased parameter
estimates even in the high-count regime, depending on the specific variance
prescription used (e.g., \citealt{Humphrey2009Chi2Bias}).
Accordingly, for Poisson-count spectra we generally recommend likelihoods that
retain the counting nature of the data (e.g., C-stat/W-stat/PG-stat) when
applicable.

\subsection{Joint likelihood}\label{sec:stats:joint}

For multiple datasets, \ELISA\ forms the joint likelihood as the product of the
per-dataset likelihoods,
\begin{equation}
  \mathcal{L}(\Theta) = \prod_{d=1}^D \mathcal{L}_d(\Theta),
\end{equation}
with shared parameters naturally inducing coupled constraints across datasets.

\subsection{Deviance}\label{sec:stats:deviance}

A useful statistic is the \emph{deviance}, which can be viewed as a rough measure of model fit
\citep[e.g.,][]{McCullagh1989GLM}.
For a likelihood $\mathcal{L}(\Theta)$, the deviance is defined as
\begin{equation}
  D(\Theta)\equiv -2\ln \mathcal{L}(\Theta) + C(\mathcal{D})
  = 2\left[\ln \mathcal{L}_{\mathrm{sat}}(\mathcal{D})
  - \ln \mathcal{L}(\Theta)\right],
\end{equation}
where $\mathcal{D}$ denotes the observed data and
$\mathcal{L}_{\mathrm{sat}}$ is the likelihood of a saturated model that fits each channel
perfectly. The additive term $C(\mathcal{D})$ depends only on the observed data
and therefore does not affect parameter inference; it is included so that
$D(\Theta)\ge 0$ and equals zero at the saturated model.

When the fitted model is correctly specified and standard regularity conditions
hold (in particular, an interior maximum and a positive-definite Fisher
information/Hessian so that parameters are locally identifiable), the deviance
evaluated at the maximum likelihood estimate, $D(\hat\Theta)$, is asymptotically
distributed as a chi-square random variable with
$\mathrm{dof}=N_{\mathrm{chan}}-N_{\mathrm{par}}$, where
$N_{\mathrm{chan}}=\sum_d J_d$ is the total number of grouped channels and
$N_{\mathrm{par}}$ is the number of free parameters \citep{Wilks1938}. This asymptotic result
motivates using $D/\mathrm{dof}\approx 1$ as a rough check, while it should be noted
that it can fail in common spectral-fitting edge cases (e.g., low counts and
boundary constraints).

For Poisson data with observation $N$ and expectation $\lambda(\Theta)$, the
per-channel Poisson deviance contribution is
\begin{equation}
  D_{\mathrm{P}}(N,\lambda)=
  2\left[N\ln\!\left(\frac{N}{\lambda}\right)-(N-\lambda)\right].
\end{equation}
For Gaussian data with observation $y$, mean $\mu(\Theta)$, and known standard
deviation $\sigma$, the deviance contribution reduces (up to an irrelevant
constant) to the familiar chi-square form,
\begin{equation}
  D_{\mathrm{G}}(y,\mu,\sigma)=\left(\frac{y-\mu}{\sigma}\right)^2.
\end{equation}

For likelihoods with an explicit background measurement (W-stat and PG-stat),
\ELISA\ forms the joint on/off likelihood and profiles out the background in
each channel to obtain $\hat b$ (Appendix~\ref{app:profile-background}). Since
the joint likelihood factorizes into on and off terms, the profile deviance is
simply the sum of the corresponding on-spectrum and off-spectrum deviance
contributions evaluated at $\hat b$: substitute the on/off data and their model
expectations into the deviance expressions above and add over channels.

\section{Posterior Inference}\label{sec:inference}

Given the likelihood $\mathcal{L}(\Theta)$ defined in Section~\ref{sec:stats},
\ELISA\ performs Bayesian inference by sampling from the posterior
\begin{equation}
  p(\Theta \mid \mathcal{D}) \propto \mathcal{L}(\Theta)\,p(\Theta),
\end{equation}
where $\mathcal{D}$ denotes all datasets and $p(\Theta)$ denotes the user-specified
(or default) prior associated with each parameter.
In \ELISA, posterior inference is primarily performed by Markov chain Monte Carlo (MCMC) samplers for parameter estimation and uncertainty quantification.
When evidence estimates are desired for model comparison, \ELISA\ also provides several nested sampling (NS; \citealt{Skilling2004NS}) alternatives.
Internally, the forward-folded model and log density are implemented in \texttt{JAX}
and exposed through a \texttt{NumPyro} probabilistic program \citep{phan2019composable},
enabling automatic differentiation and execution on CPU, GPU, or TPU.
This \texttt{NumPyro} program serves as the canonical JAX implementation of the
log posterior; external inference backends interface with \ELISA\ through the same
JAX-compiled log-likelihood/log-posterior callbacks.

\subsection{Sampling backends}\label{sec:inference:samplers}

Currently, \ELISA\ supports sampling from the posterior with several third-party inference engines, which can be categorized into (i) gradient-based MCMC, (ii) gradient-free MCMC, and (iii) nested sampling for evidence estimation.
They can be used with \ELISA\ through a consistent interface\footnote{\url{https://astro-elisa.readthedocs.io/en/stable/apidoc/elisa.infer.fit.html\#elisa.infer.fit.BayesFit}}.

\subsubsection{Gradient-based MCMC}\label{sec:inference:samplers:grad}

\ELISA\ supports the No-U-Turn Sampler (NUTS; \citealt{Hoffman2014NUTS}) via
\texttt{NumPyro} and \texttt{BlackJAX} \citep{cabezas2024blackjax}.
NUTS is an adaptive Hamiltonian Monte Carlo (HMC; see e.g., \citealt{Neal2011HMC,Betancourt2013HMC}) method that eliminates manual choice of the trajectory length by terminating proposals when the simulated path starts to retrace itself.
By exploiting gradients of the log posterior, NUTS can generate distant proposals with reduced random-walk behavior and often achieves high effective sample sizes per gradient evaluation for smooth posteriors in moderate to high dimensions.
Its limitations are that it requires a differentiable model and can be sensitive to challenging posterior geometry; backend-reported warnings such as divergent transitions or related integration/adaptation pathologies should therefore be checked routinely and, when persistent, typically motivate reparameterization, stronger weakly informative priors, or more conservative adaptation settings.

For cases where full Hamiltonian trajectories are unnecessary or unstable, \ELISA\ also provides a gradient-based Metropolis--Hastings alternative based on the Barker proposal \citep{Livingstone2022BarkerProposal} via \texttt{NumPyro}.
Compared to NUTS, it can be simpler to configure and can be competitive in low to moderate dimensions, but it typically mixes more slowly for strongly correlated or high-dimensional posteriors and may require more likelihood evaluations per effective draw.

\subsubsection{Gradient-free MCMC}\label{sec:inference:samplers:gradfree}

When gradients are unavailable (e.g., genuinely non-differentiable components) or
unreliable, \ELISA\ provides gradient-free alternatives.

\ELISA\ supports ensemble-based MCMC using both \texttt{JAX}-native implementations and
optional external backends. Through \texttt{NumPyro}, it provides
affine-invariant ensemble sampling \citep{Goodman2010AIES, Foreman-Mackey13emcee} and ensemble
slice sampling \citep{Karamanis2020ESS, Karamanis2021zeus} in \texttt{JAX} native implementations. These methods are gradient-free and can be effective for correlated posteriors, but they typically require many walkers, increasing the likelihood-evaluation budget for expensive forward-folded models.
As a practical guideline, ensemble samplers generally benefit from using a number of walkers at least of order twice the parameter dimension.
Like most MCMC methods, ensemble samplers may struggle to move between well-separated modes without careful initialization, so walker-level mixing diagnostics and problem-specific initialization strategies are important.
\ELISA\ exposes user-specified initial values through its fitting interface, and sampler outputs are converted to \texttt{ArviZ} \texttt{InferenceData} objects so that initialization choices, trace diagnostics, and backend-specific warnings can be inspected in a consistent way.

Optional wrappers to \texttt{emcee} \citep{Foreman-Mackey13emcee} and \texttt{zeus} \citep{Karamanis2021zeus} are provided for compatibility and convenient cross-checks for ensemble samplers mentioned above.
Compared to the \texttt{JAX}-native ensemble methods, wrappers to external backends may incur additional overhead and are therefore most practical for low-dimensional or moderate-cost likelihoods, or when reproducing legacy workflows.

In addition, \ELISA\ provides Sample Adaptive MCMC (SA; \citealt{Zhu2019SAMCMC})
via \texttt{NumPyro} as another gradient-free option. SA can offer a robust,
largely automated alternative in moderate dimensions, but it typically requires
a sufficiently long adaptation phase, and its efficiency can degrade for
strongly correlated posteriors.

\subsubsection{Nested sampling}\label{sec:inference:samplers:ns}

Nested sampling targets the marginal likelihood (evidence) $Z=\int \mathcal{L}(\Theta)p(\Theta)\,\mathrm{d}\Theta$ while producing posterior samples as a byproduct, enabling evidence-based model comparison.
It can be advantageous for multimodal posteriors or for problems where evidence estimation is required, but it is typically more computationally demanding than MCMC and becomes challenging as the parameter dimension grows.
\ELISA\ supports multiple nested sampling backends, including \texttt{UltraNest} \citep{Buchner2016UltraNest,Buchner2019UltraNest,Buchner2021UltraNest}, \texttt{Nautilus} \citep{Lange2023NAUTILUS}, and \texttt{JAXNS} \citep{Albert2020JAXNS, Albert2023JAXNS}, which
return both posterior samples and an evidence estimate (with an uncertainty estimate when available).
Among these, \texttt{UltraNest} provides a robust general-purpose nested sampling engine with well-tested evidence estimation and practical run-time diagnostics.
\texttt{Nautilus} implements importance nested sampling boosted by deep learning, which can reduce the number of likelihood evaluations for challenging posteriors at the cost of additional algorithmic overhead.
\texttt{JAXNS} provides a \texttt{JAX}-native nested sampling implementation and supports modern variants such as phantom-powered nested sampling \citep{Albert2023JAXNS}, making it well suited to accelerated, vectorized likelihood evaluation.

\subsection{Convergence diagnostics}\label{sec:inference:convergence}
Posterior summaries (e.g., uncertainty intervals and derived quantities) are only meaningful after the sampler has reached the stationary distribution.
Accordingly, \ELISA\ treats convergence assessment as a routine step of the Bayesian workflow.
To support a standardized diagnostic workflow across inference backends, \ELISA\ relies on \texttt{ArviZ} \citep{Kumar2019Arviz}, a \texttt{Python} package for exploratory analysis of Bayesian models.
\ELISA\ packages posterior samples, pointwise log likelihoods, and sampler diagnostics into a \texttt{InferenceData} object of \texttt{ArviZ}, and then uses its downstream utilities to compute and visualize convergence metrics.

\subsubsection{Core diagnostics for MCMC samplers}
For MCMC sampler, a two-stage procedure is recommended: visual checks for stationarity and mixing
followed by quantitative diagnostics of convergence and Monte Carlo accuracy.
Trace plots provide a first-pass assessment of stationarity and mixing.

For quantitative convergence checks, \ELISA\ reports the rank-normalized split-$\widehat{R}$ diagnostic and bulk effective sample sizes \citep{Vehtari2021Rhat} after posterior sampling.
As a practical rule of thumb, it is recommended to target $\widehat{R}\lesssim 1.01$ for all reported parameters and key derived quantities, while noting that $\widehat{R}\approx 1$ is necessary but not sufficient for reliable inference.
Since there are autocorrelations in the chains, it is also recommended to inspect both bulk and tail effective sample sizes (ESS) to assess Monte Carlo accuracy, which are relevant for point estimates and tail-sensitive summaries, respectively \citep{Vehtari2021Rhat}. A practical guideline is to target ESS $\gtrsim 400$ for all reported parameters and derived quantities.

\subsubsection{Sampler-specific diagnostics and remedies}
For Hamiltonian methods (NUTS/HMC), additional diagnostics include divergent
transitions, saturation of the maximum tree depth, and energy-related warnings;
persistent issues typically indicate problematic posterior geometry or overly
aggressive integration settings, and are often mitigated by reparameterization,
strengthening weakly informative priors, or using more conservative adaptation
targets (e.g., higher target acceptance probability).
For ensemble samplers, mixing should be assessed by verifying cross-walker
movement across the typical set and by monitoring autocorrelation lengths and
acceptance behavior. For nested sampling, convergence is assessed through the
reported uncertainty on $\ln Z$ and the effective number of posterior samples;
insufficient accuracy can usually be addressed by increasing the live-point
budget or tightening termination criteria.

\subsection{Example: \textit{Insight}-HXMT Crab observation}

\begin{figure*}[t]
  \centering
  \includegraphics[width=0.7\textwidth]{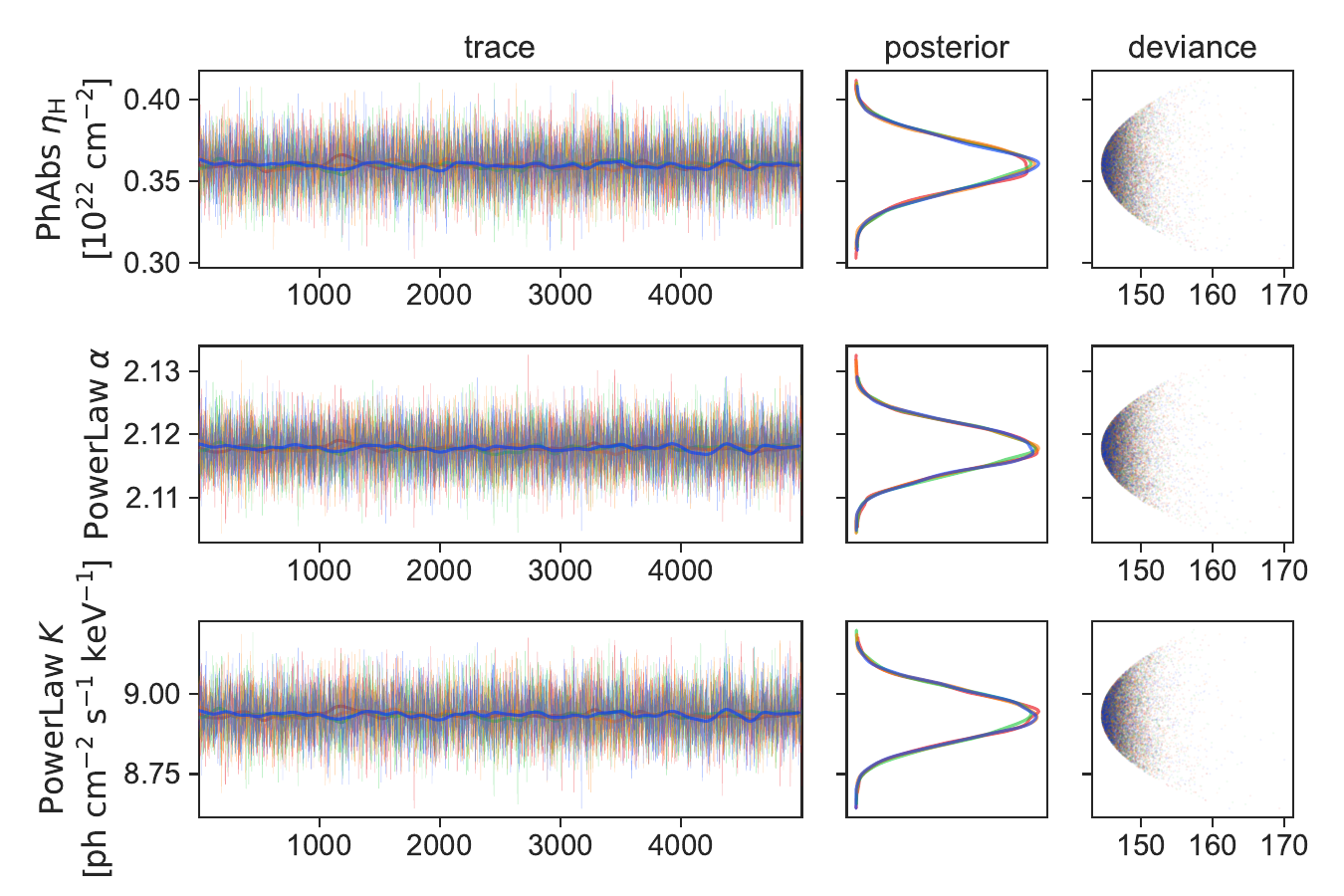}
  \caption{
    Convergence diagnostics for the HXMT Crab example.
    For each parameter, the left panel shows the post-warm-up trace for the four chains (distinguished by color), together with a dark-blue smoothed running mean as a visual guide; the middle panel shows the marginal posterior density by chain, and the right panel shows the per-draw deviance cloud.
    Stable traces, overlapping densities, and the absence of systematic drifts or isolated high-deviance excursions indicate good mixing and convergence.}
  \label{fig:conv-diagnostics}
\end{figure*}

Here we illustrate the MCMC sampling diagnostics on a joint fit to the Low Energy (LE) Telescope, Medium Energy (ME) Telescope, and High Energy (HE) Telescope spectra from a pointed \textit{Insight}-HXMT observation of the Crab (ObsID P011160506306; \citealt{Zhang20HXMTOverview}).
We restrict the analysis to 1.5--10~keV (LE), 10--35~keV (ME), and 28--250~keV (HE), and apply response-aware optimal binning \citep{Kaastra2016OptBin} to the datasets before fitting, resulting in 131 grouped fitted bins across the three instruments.
The model is an absorbed power law (PhAbs $\times$ PowerLaw), and each dataset uses the PG-stat likelihood because the source spectrum is treated as Poisson counts while the associated background product is represented by a Gaussian estimate with its own uncertainty within the standard HXMT workflow (Section~\ref{sec:stats:joint}).

We first perform posterior sampling with NUTS using 4 chains, 2000 warm-up iterations, and 5000 post-warm-up draws per chain.
As shown in Figure~\ref{fig:conv-diagnostics}, the chains mix well for all three parameters, the chain-wise marginal posteriors closely overlap, and the deviance scatter plot shows no outliers of high-deviance values.
For this fit, \ELISA\ reports rank-normalized split-$\widehat{R}=1.00$ and bulk ESS of $\sim10^4$ for the three parameters.

\begin{figure}[t]
  \centering
  \includegraphics[width=\columnwidth]{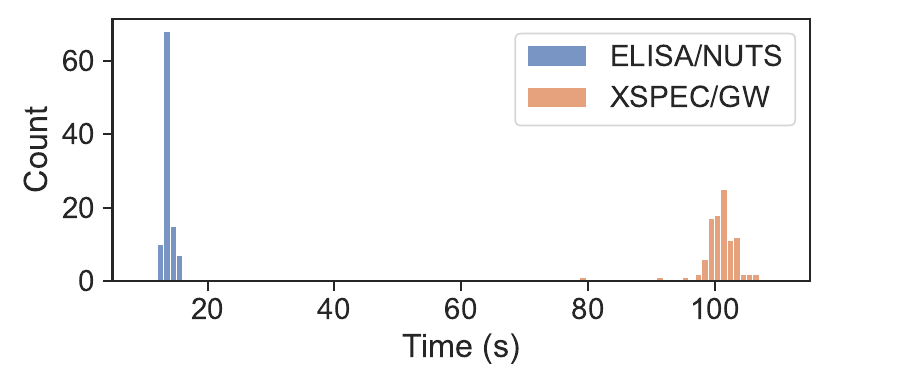}
  \includegraphics[width=\columnwidth]{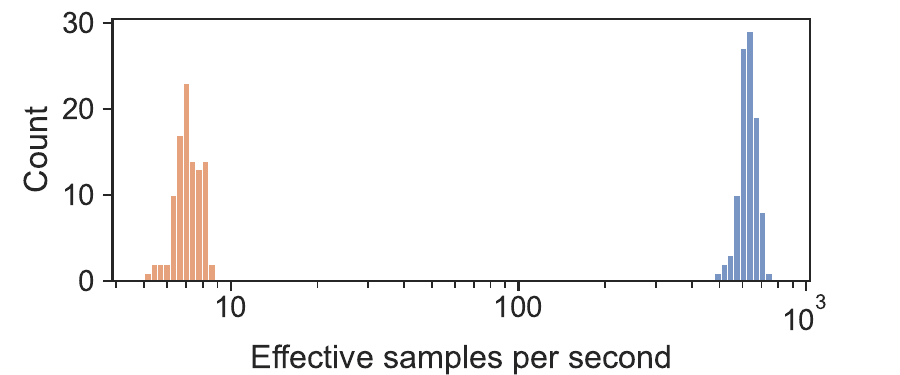}
  \includegraphics[width=\columnwidth]{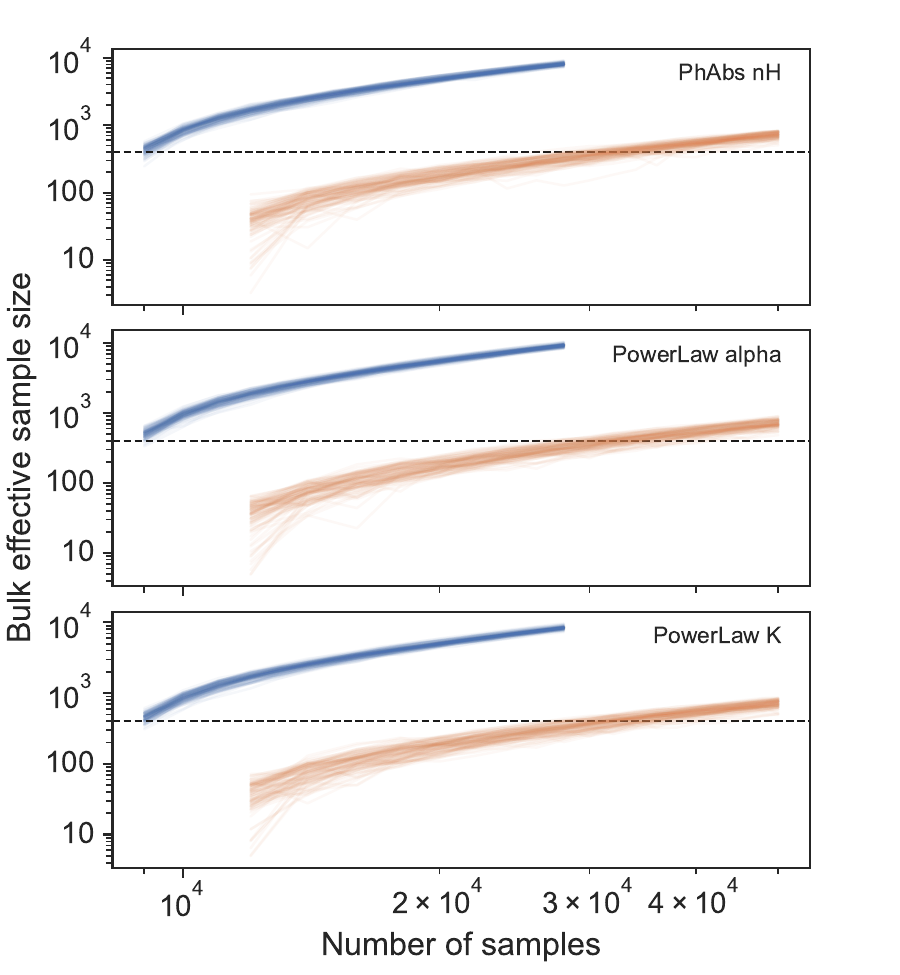}
  \caption{Sampling efficiency diagnostics for the HXMT Crab example.
  \textbf{Top}: distribution of wall-clock sampling times over 100 repeated fits for \ELISA/NUTS (blue) and \XSPEC/GW (orange).
  \textbf{Middle}: distribution of effective samples per second of each fit for \ELISA/NUTS (blue) and \XSPEC/GW (orange).
  \textbf{Bottom}: bulk ESS as a function of total sample count for the same repeats (one thin line per run); the dashed line marks $\mathrm{ESS}_{\mathrm{bulk}}=400$.}
  \label{fig:inference-time-ess}
\end{figure}

To characterize runtime variability and provide a configuration-specific baseline comparison rather than a universal benchmark, we repeated the same fit 100 times with different random seeds and recorded the total wall-clock sampling time and bulk ESS, using four Apple M1 Max CPU devices. In this example, both \ELISA\ and \XSPEC\ used the same absorbed power-law spectral parameterization, with matched MCMC settings, including the same number of chains, burn-in steps, and post-burn-in sampling steps. The comparison is therefore intended to assess posterior-sampling efficiency under the explicitly stated software settings and hardware configuration, rather than to establish a general performance ranking between the two systems. Figure~\ref{fig:inference-time-ess} shows that \ELISA/NUTS typically completes in $\sim$12--16~s per fit, whereas an \XSPEC\ Goodman--Weare ensemble sampler (\XSPEC/GW) configured with 8 walkers, 10{,}000 burn-in steps, and 40{,}000 retained steps requires $\sim$80--110~s.
At the end of each run, \ELISA/NUTS attains bulk ESS $\sim 10^4$ for these parameters, while the ensemble chain yields bulk ESS $\lesssim 10^3$.
In terms of effective samples per unit time, this corresponds to a $\sim 100\times$ improvement in sampling efficiency for this example, showing that by leveraging JAX-accelerated, differentiable forward folding, Bayesian posterior sampling could be much more efficient. \cite{Dupourque2024jaxspec} has shown that, NUTS also converges much faster than the ensemble sampler, when using rank-normalized split-$\widehat{R}\lesssim 1.01$ as the convergence diagnostic.

\section{Model Validation and Comparison}\label{sec:diagnostics}

After inference (Section~\ref{sec:inference}), a forward-folded spectral analysis
must address two distinct questions: (i) whether the fitted model provides an
adequate description of the data under the adopted likelihood, and (ii) which
model among plausible alternatives offers better predictive performance. In
\ELISA, model checking and comparison are designed to be part of the iterative
analysis loop (Figure~\ref{fig:workflow}), motivating revisions to model
structure, priors, or likelihood assumptions when diagnostics indicate
systematic tension.

\subsection{Model validation}

\subsubsection{Deviance per degree of freedom}\label{sec:diagnostics:stat-dof}

As a quick sanity check, many workflows report a single reduced statistic (e.g.,
reduced $\chi^2$). \ELISA\ similarly reports the total deviance $D$ together with
a nominal number of degrees of freedom (Section~\ref{sec:stats:deviance}).
This summary is useful for flagging gross pathologies (e.g., an obviously wrong
model family or response usage), but it is not a calibrated measure of fit for
typical counting spectra: the reference distribution of $D$ can deviate
substantially from a $\chi^2$ approximation, and the notion of $\mathrm{dof}$
becomes ambiguous under strong priors, non-identifiability, or multi-dataset
parameter tying. For substantive model criticism, \ELISA\ therefore emphasizes diagnostics described below, using $D/\mathrm{dof}$ only
as a coarse screen.

\subsubsection{Leave-one-out predictive calibration (PSIS-LOO): PIT, Q--Q, and quantile residuals}
\label{sec:diagnostics:loo}

A well-specified forward model should not only match the observed counts
pointwise, but also yield \emph{distributionally calibrated} predictions for
each channel. Conceptually, calibration is most cleanly assessed using an
out-of-sample predictive distribution. \ELISA\ adopts the leave-one-out (LOO)
predictive view: each channel is checked against the predictive distribution
conditioned on all \emph{other} channels, thereby reducing the ``double use'' of
the same data unit for both fitting and checking.

Let $y_j$ denote the channel-wise data unit (for background-inclusive
likelihoods, $y_j$ includes the associated background term(s) for that channel),
and let $F_j(y\mid\Theta)\equiv \Pr(Y^{\mathrm{rep}}_j\le y\mid\Theta)$ be the
predictive CDF under parameters $\Theta$, where $Y^{\mathrm{rep}}_j$ denotes a replicated version of the $j$-th data unit drawn from the fitted predictive distribution. Here PIT denotes the probability integral transform. The LOO predictive CDF is
\begin{equation}
  F_{\mathrm{loo},j}(y)\equiv \Pr(Y^{\mathrm{rep}}_j\le y\mid \mathcal{D}_{-j})
  = \int F_j(y\mid\Theta)\,p(\Theta\mid\mathcal{D}_{-j})\,d\Theta ,
\end{equation}
where $\mathcal{D}_{-j}$ removes the $j$-th data unit.
The LOO-PIT value is $u_{\mathrm{loo},j}\equiv F_{\mathrm{loo},j}(y_j)$ (or its
randomized form for discrete outcomes; see below). Under correct specification,
$\{u_{\mathrm{loo},j}\}$ should be approximately $\mathrm{Uniform}(0,1)$; the
corresponding LOO quantile residual
\begin{equation}
  r_{\mathrm{loo},j}\equiv \Phi^{-1}(u_{\mathrm{loo},j})
\end{equation}
should be approximately $\mathcal{N}(0,1)$, where $\Phi$ is the standard normal cumulative distribution function, enabling quantile--quantile (Q--Q) and residual plots
with a common reference scale across likelihood choices.

\paragraph{Discrete outcomes and randomized PIT.}
For discrete likelihoods, $F_{\mathrm{loo},j}$ is stepwise and the PIT is
interval-valued. Following \citealt{Dunn1996RQR,Warton17PITtrap}, \ELISA\ uses a
randomized PIT. For Poisson data,
\begin{equation}
  u_{\mathrm{loo}} \sim
  \mathrm{Uniform}\!\left(F_{\mathrm{loo}}(N-1),\,F_{\mathrm{loo}}(N)\right),
  \qquad
  r_{\mathrm{loo}}=\Phi^{-1}(u_{\mathrm{loo}}).
\end{equation}

For likelihoods that profile the background (Appendix~\ref{app:profile-background}),
\ELISA\ evaluates calibration on a scalar net statistic
$X\equiv N-\alpha N^{\mathrm{off}}$ implied by the on/off formulation. Let $s$
be the source expectation, $\alpha$ the on/off scaling factor, and $\hat b(s)$
the profiled background solution. For W-stat (Poisson on/off),
\begin{equation}
  \begin{aligned}
    N^{\mathrm{rep}} &\sim \mathrm{Poisson}(s+\alpha \hat b),\\
    N^{\mathrm{off,rep}} &\sim \mathrm{Poisson}(\hat b),\\
    X^{\mathrm{rep}} &\equiv N^{\mathrm{rep}}-\alpha N^{\mathrm{off,rep}}.
  \end{aligned}
\end{equation}
Because $X$ is discrete (and not necessarily integer-valued when $\alpha$ is not
an integer), the PIT is again interval-valued. With
$x\equiv N-\alpha N^{\mathrm{off}}$ and the next smaller attainable value
\begin{equation}
  x^{-}\equiv \max\!\Big((N-1)-\alpha N^{\mathrm{off}},\; N-\alpha(N^{\mathrm{off}}+1)\Big),
\end{equation}
the randomized PIT uses
\begin{equation}
  u_{\mathrm{loo}} \sim \mathrm{Uniform}\!\left(F_{\mathrm{loo},X}(x^{-}),\,F_{\mathrm{loo},X}(x)\right),
\end{equation}
where $F_{\mathrm{loo},X}$ is the LOO predictive CDF of $X^{\mathrm{rep}}$.

For PG-stat (Poisson on with Gaussian background estimate),
\begin{equation}
  \begin{aligned}
    N^{\mathrm{rep}} &\sim \mathrm{Poisson}(s+\alpha \hat b),\\
    N^{\mathrm{off,rep}} &\sim \mathcal{N}(\hat b,\sigma^{\mathrm{off}}),\\
    X^{\mathrm{rep}} &\equiv N^{\mathrm{rep}}-\alpha N^{\mathrm{off,rep}},
  \end{aligned}
\end{equation}
so $X$ is continuous and $u_{\mathrm{loo}}=F_{\mathrm{loo},X}(x)$ without
randomization. In practice, \ELISA\ evaluates $F_{\mathrm{loo},X}$ by Monte Carlo
predictive replicates, e.g.,
$F_{\mathrm{loo},X}(x)\approx R^{-1}\sum_{r=1}^R\mathbb{I}(X^{(r)}\le x)$,
where $R$ is the number of predictive replicates.

Exact LOO would require refitting the model $J$ times. \ELISA\ instead
approximates $p(\Theta\mid\mathcal{D}_{-j})$ from posterior draws
$\{\Theta^{(s)}\}_{s=1}^S\sim p(\Theta\mid\mathcal{D})$ using Pareto-smoothed
importance sampling (PSIS; \citealt{Vehtari2015LOO}). For each $j$,
\begin{equation}
  F_{\mathrm{loo},j}(y)\approx
  \sum_{s=1}^S \tilde w^{(s)}_j\,F_j(y\mid\Theta^{(s)}),
  \qquad
  \sum_{s=1}^S \tilde w^{(s)}_j = 1,
\end{equation}
with raw importance ratios $w^{(s)}_j\propto 1/p(y_j\mid\Theta^{(s)})$.
PSIS stabilizes heavy-tailed ratios by smoothing and truncating the largest
weights; the fitted generalized-Pareto tail-shape parameter (reported as
Pareto-$k$) provides a practical reliability diagnostic. In \ELISA,
$k\lesssim 0.7$ typically indicates stable reweighting, while larger values
flag potentially unreliable approximations and motivate remedial actions
(e.g., stronger regularization, reparameterization, or exact refits for the
affected channels). The same PSIS-LOO machinery underlies the predictive
criteria used for model comparison (Section~\ref{sec:diagnostics:compare}).

\begin{figure}[t]
	\centering
	\includegraphics[width=0.42\textwidth]{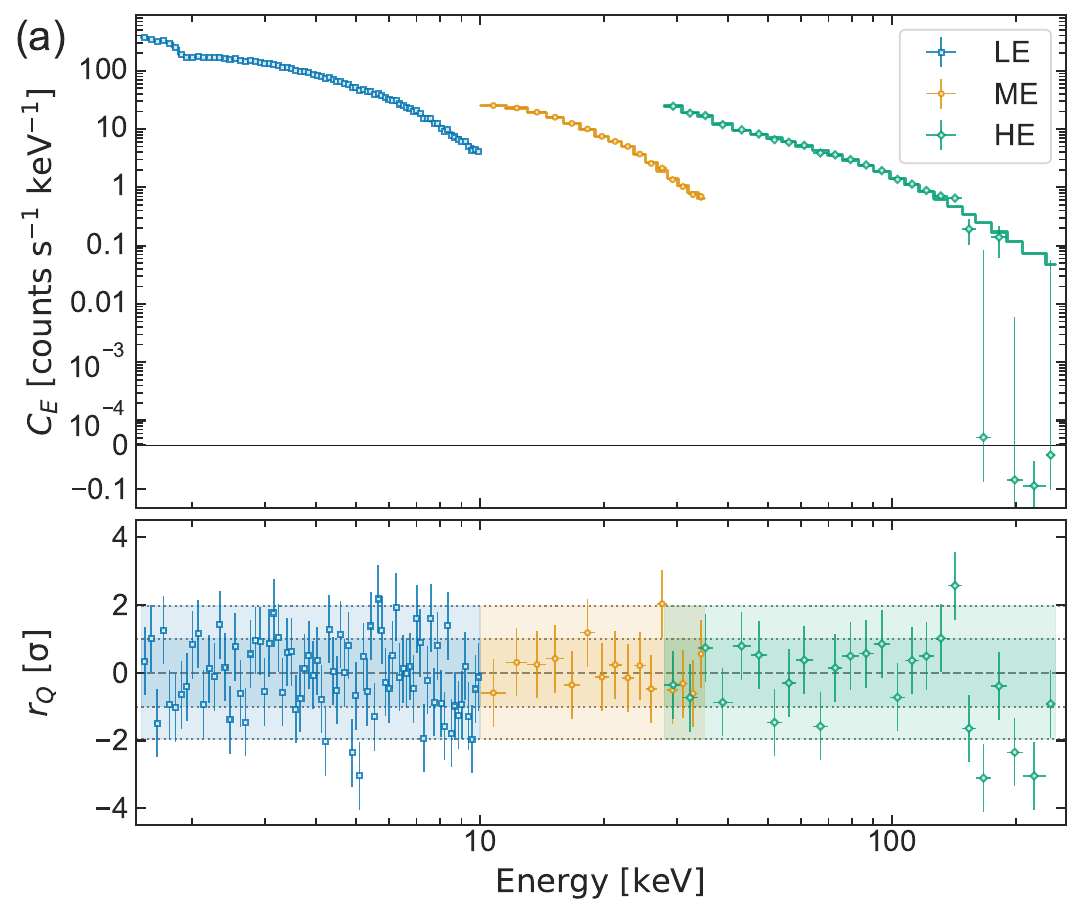}
  \includegraphics[width=0.42\textwidth]{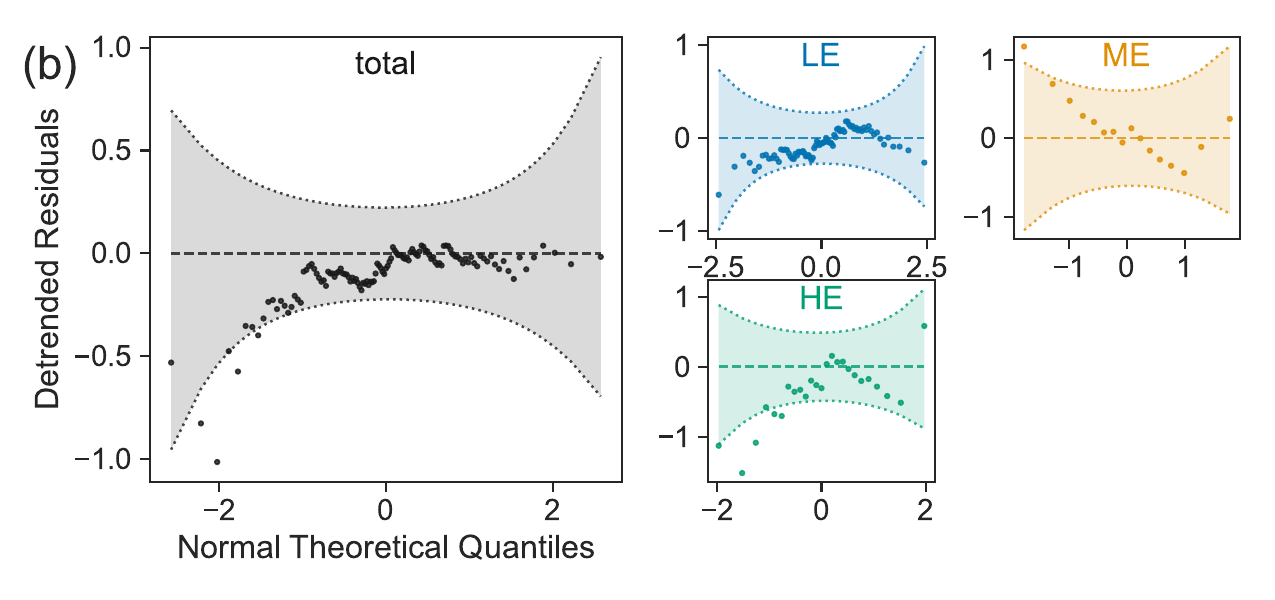}
  \includegraphics[width=0.42\textwidth]{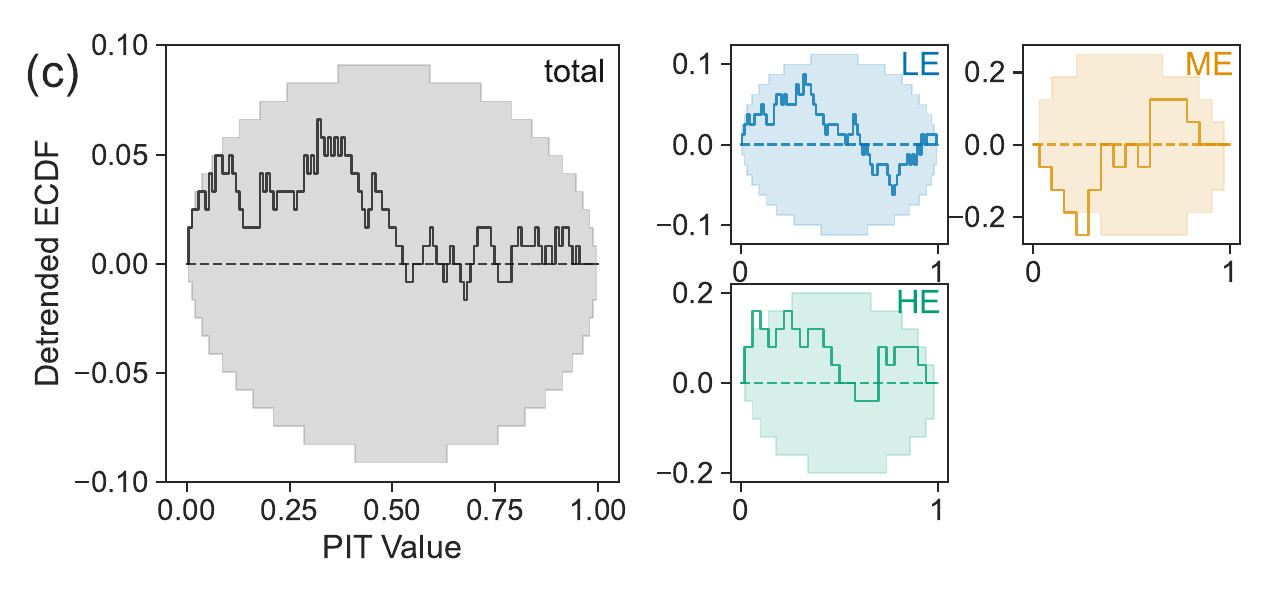}
  \includegraphics[width=0.42\textwidth]{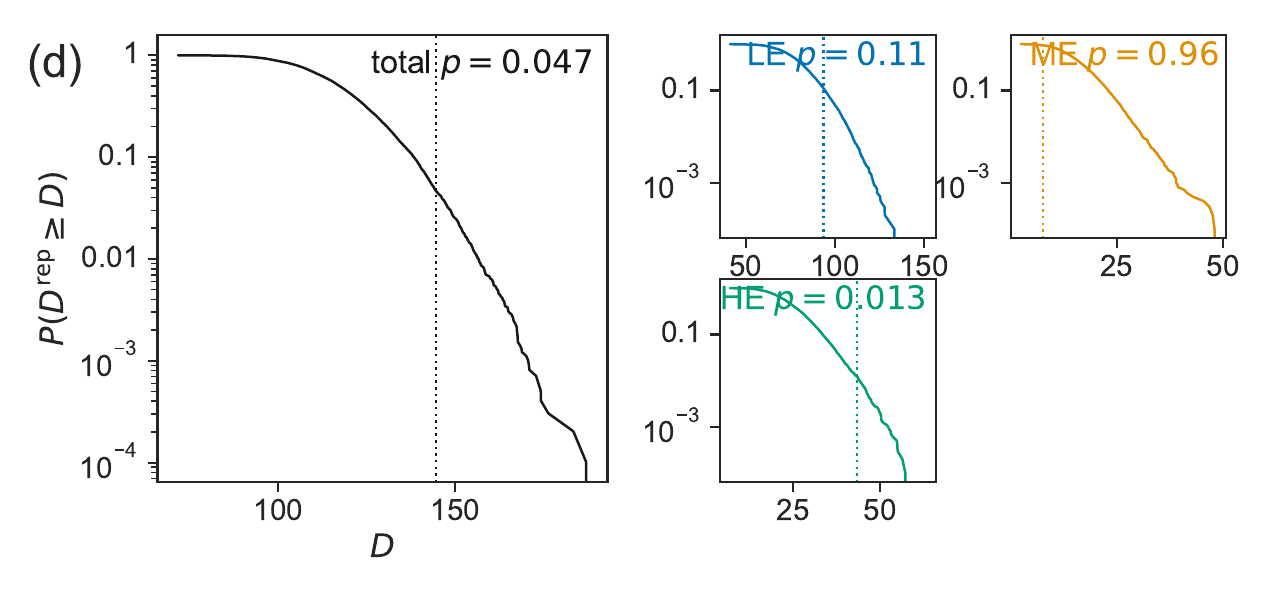}
  \caption{Example of spectral fit and residual diagnostics for the HXMT Crab observation (a), and corresponding PSIS-LOO Q--Q (b), PIT (c), and PPC (d) diagnostics.
  The shaded area around the reference line in the Q--Q (b) and PIT (c) plots represents the 95\% pointwise confidence interval. Systematic deviations outside the confidence interval highlight
  miscalibration. The detrending removes the linear trend, making it easier to see systematic deviations. The PPC panel (d) compares the observed deviance to its posterior predictive distribution from simulated-and-refit replicates and reports a posterior predictive $p$-value; values near 0 or 1 indicate tension between the data and the fitted model.}
  \label{fig:loo-diagnostics}
\end{figure}

To obtain residuals with a common interpretation across likelihoods, \ELISA\
uses PIT-based quantile residuals, with the LOO variant
$r_{\mathrm{loo},j}=\Phi^{-1}(u_{\mathrm{loo},j})$ as the default residual. These
residuals support both energy-localized inspection (as curves) and global
calibration assessment (via their Q--Q and PIT summaries).
Figure~\ref{fig:loo-diagnostics} illustrates PSIS-LOO calibration diagnostics
for the \textit{Insight}-HXMT Crab example in
Section~\ref{sec:inference:convergence}. Departures from the PIT reference line
are often diagnostically informative: a monotone shift indicates systematic
bias, symmetric S-shaped deviations indicate miscalibrated predictive spread
(under- or overdispersion), and asymmetric deviations suggest skewness or tail
mismatch. For a more detailed taxonomy of PIT ECDF patterns and corresponding
failure modes, see \cite{Sailynoja22PITECDF}.

\subsubsection{Posterior predictive checks and \texorpdfstring{$p$}{p}-values}
\label{sec:diagnostics:ppc}

Residual diagnostics and LOO-PIT focus on calibration at the level of
channel-wise data units. For a complementary global assessment of adequacy,
\ELISA\ provides posterior predictive checks (PPC; \citealt{Gelman1996PPC}),
comparing the observed dataset to replicated datasets simulated from the fitted
generative model (including the adopted background treatment).
Given posterior draws $\{\Theta^{(s)}\}$, \ELISA\ generates replicated spectra
$\mathcal{D}^{\mathrm{rep},(s)}\sim p(\mathcal{D}\mid \Theta^{(s)})$ from the
same likelihood used for fitting and computes a discrepancy statistic
$T(\mathcal{D})$. In the current implementation, $T$ is based on the fitted
deviance (Section~\ref{sec:stats:deviance}): for each replicate we refit the
model using the same fitting procedure as for the observed data to obtain
$\hat\Theta^{\mathrm{rep},(s)}$, and set
\begin{equation}
  T\!\left(\mathcal{D}^{\mathrm{rep},(s)}\right)
  \equiv D\!\left(\mathcal{D}^{\mathrm{rep},(s)};\hat\Theta^{\mathrm{rep},(s)}\right),
  \qquad
  T(\mathcal{D})\equiv D(\mathcal{D};\hat\Theta).
\end{equation}
This yields a posterior predictive distribution of fitted deviances, and the
posterior predictive $p$-value is
\begin{equation}
  p_{\mathrm{PPC}} \equiv
  \Pr\!\left[T\!\left(\mathcal{D}^{\mathrm{rep}}\right)\ge T(\mathcal{D})\,\middle|\,\mathcal{D}\right],
\end{equation}
which we approximate by Monte Carlo as
\begin{equation}
  \hat p_{\mathrm{PPC}} =
  \frac{1}{S}\sum_{s=1}^{S}\mathbb{I}\!\left[T\!\left(\mathcal{D}^{\mathrm{rep},(s)}\right)\ge T(\mathcal{D})\right].
\end{equation}
reported per dataset and for the joint fit. Values near 0 or 1 indicate tension
between the observed data and the posterior predictive distribution. As with
other PPC diagnostics, these $p$-values are calibrated measures of discrepancy
under the fitted model rather than uniform frequentist $p$-values.

\subsection{Model comparison}\label{sec:diagnostics:compare}

\ELISA\ supports both predictive and evidence-based approaches to comparing
candidate models. Predictive approaches target expected out-of-sample accuracy
and are typically more robust to prior choices; evidence-based approaches
compare marginal likelihoods when these can be estimated reliably.

\subsubsection{PSIS-LOO and WAIC}

\ELISA\ computes PSIS-LOO \citep{Vehtari2015LOO} and WAIC
\citep{Watanabe2010WAIC} from the pointwise log likelihood stored during
sampling. Both estimate expected out-of-sample predictive performance and
provide uncertainty estimates and effective parameter counts. In \ELISA, they
are reported on the deviance scale ($-2\times$ expected log predictive density),
so smaller values indicate better expected predictive performance.

For a joint fit, let $\mathcal{D}=\{y_j\}_{j=1}^{J}$ denote channel-wise data
units across all datasets (including any background terms required by the chosen
likelihood; Section~\ref{sec:stats}). The LOO expected log predictive density is
\begin{equation}
  \mathrm{elpd}_{\mathrm{loo}} \equiv \sum_{j=1}^{J} \ln p(y_j\mid \mathcal{D}_{-j}),
  \qquad
  \mathrm{LOOIC}\equiv -2\,\mathrm{elpd}_{\mathrm{loo}} ,
\end{equation}
with $p(y_j\mid\mathcal{D}_{-j})$ approximated via PSIS reweighting as described
in Section~\ref{sec:diagnostics:loo}. \ELISA\ reports Pareto-$k$ diagnostics to
assess the stability of this approximation.

WAIC is computed from posterior draws $\{\Theta^{(s)}\}_{s=1}^{S}$ using the
pointwise log likelihood values $\ln p(y_j\mid\Theta^{(s)})$:
\begin{equation}
  \begin{split}
    \mathrm{elpd}_{\mathrm{waic}}
    \equiv {} &
    \sum_{j=1}^{J}\ln\!\left(\frac{1}{S}\sum_{s=1}^{S} p(y_j\mid\Theta^{(s)})\right) \\
    &- \sum_{j=1}^{J}\mathrm{Var}_{s}\!\left[\ln p(y_j\mid\Theta^{(s)})\right],
  \end{split}
\end{equation}
\begin{equation}
  \mathrm{WAIC}\equiv -2\,\mathrm{elpd}_{\mathrm{waic}} .
\end{equation}

In practice, PSIS-LOO is preferred when Pareto-$k$ indicates stable reweighting \citep{Vehtari2015LOO}.
For model comparison, differences in LOOIC/WAIC should be interpreted together
with their standard errors: differences small compared to the uncertainty do
not support a decisive preference.

\subsubsection{Evidence and Bayes factors}

When nested sampling is used (Section~\ref{sec:inference:samplers:ns}), \ELISA\
records an estimate of the marginal likelihood $Z$ (and its uncertainty),
enabling Bayes-factor comparisons between models. For two candidate models
$M_1$ and $M_2$ fitted to the same dataset $\mathcal{D}$, the Bayes factor is
\begin{equation}
  B_{12} \equiv \frac{p(\mathcal{D}\mid M_1)}{p(\mathcal{D}\mid M_2)}
  = \frac{Z_1}{Z_2},
  \qquad
  \ln B_{12} = \ln Z_1 - \ln Z_2 ,
\end{equation}
and updates prior odds to posterior odds via \citep{Kass1995BayesFactors}
\begin{equation}
  \frac{p(M_1\mid \mathcal{D})}{p(M_2\mid \mathcal{D})}
  = B_{12}\,\frac{p(M_1)}{p(M_2)} .
\end{equation}
Because evidence ratios integrate over the prior, they naturally encode an
Occam penalty for unused prior volume, but they can be sensitive to prior choice
and parameterization. Accordingly, evidence-based conclusions should be
reported together with the corresponding priors and, when feasible, sensitivity
analyses that assess robustness to reasonable prior variations. This is
particularly important when parameters are weakly identified (e.g., a cutoff
energy far above the observed range), where Bayes factors can be driven
primarily by assumed prior volume rather than by the likelihood.

\subsubsection{Model averaging}

In many practical analyses, model comparison does not yield a decisive winner,
and conditioning on a single selected model can understate uncertainty. A
natural alternative is to propagate model uncertainty by averaging inferences
or predictions over a set of plausible models. In Bayesian model averaging
(BMA; \citealt{Hoeting1999BMA}), posterior model weights are
\begin{equation}
  p(M_m\mid\mathcal{D})
  = \frac{Z_m\,p(M_m)}{\sum_{k} Z_k\,p(M_k)} ,
\end{equation}
and the model-averaged posterior predictive distribution is
\begin{equation}
  p(\tilde y\mid\mathcal{D})
  = \sum_{m} p(\tilde y\mid\mathcal{D},M_m)\,p(M_m\mid\mathcal{D}) .
\end{equation}
The same weighting applies to derived quantities, e.g.,
$\mathbb{E}[g\mid\mathcal{D}]=\sum_m \mathbb{E}[g\mid\mathcal{D},M_m]\,p(M_m\mid\mathcal{D})$.
While BMA is coherent with the Bayesian evidence, it inherits the same prior
sensitivity as Bayes factors. When the primary goal is predictive accuracy,
LOO-based averaging via stacking \citep{Yao2018Stacking} can be attractive: it
chooses weights to optimize out-of-sample predictive performance using the same
leave-one-out predictive quantities underlying PSIS-LOO
(Section~\ref{sec:diagnostics:loo}).
\ELISA\ outputs the ingredients required for both approaches (evidence estimates
from nested sampling and pointwise log likelihoods for PSIS-LOO), so model
averaging can be carried out in post-processing when appropriate.

\subsubsection{Example: HXMT Crab continuum comparison}

As a concrete illustration, we consider the HXMT Crab data in
Section~\ref{sec:inference:convergence} and compare an absorbed power law
(PhAbs $\times$ PowerLaw) to an absorbed cutoff power law (PhAbs $\times$
CutoffPL). For these data, the additional cutoff parameter is typically weakly
constrained; consequently, predictive criteria (PSIS-LOO/WAIC) often show no
substantial improvement over the simpler model once uncertainty is accounted
for. Evidence-based comparisons in such cases should be interpreted with care
and accompanied by explicit prior reporting for the cutoff parameter.

Table~\ref{tab:case-study-compare} summarizes the model comparison results for
this case study, reporting deviance-based information criteria, predictive
criteria (PSIS-LOO/WAIC), and nested-sampling evidence estimates (when
available), together with uncertainty estimates and normalized weights.

\begin{deluxetable*}{lcccccccccc}
\tablecaption{Model comparison summaries ($\Delta$IC) for the HXMT Crab continuum case study.\label{tab:case-study-compare}}
\tablehead{
\colhead{Model} &
\colhead{Mean/Median Deviance} &
\colhead{$k$} &
\colhead{edof} &
\colhead{$\Delta$AIC} &
\colhead{$\Delta$BIC} &
\colhead{$\Delta$WAIC (SE)} &
\colhead{$\Delta$LOO (SE)} &
\colhead{$\Delta\ln Z$} &
\colhead{Weight}
}
\startdata
$\mathrm{PhAbs}\times\mathrm{PowerLaw}$  & 147.5/146.8 & 3 & 2.4 & / & / & / & / & / & 1.00 \\
$\mathrm{PhAbs}\times\mathrm{CutoffPL}$ & 148.4/147.8 & 4 & 2.6 & 2.37 & 5.03 & 1.19 (1.38) & 1.20 (1.38) & -2.53 & 0.00 \\
\enddata
\tablecomments{Deviance shows mean/median of posterior deviance. $k$ is the number of free parameters, and edof is the effective number of parameters obtained from LOOIC and WAIC. $\Delta$IC values are computed relative to the smallest value in this set (so the best model has $\Delta$IC$=0$). WAIC and PSIS-LOO are reported on the deviance scale ($-2\times$ expected log predictive density); values in parentheses give the standard errors of the differences. $\Delta\ln Z$ is the difference in log marginal likelihood estimated from \texttt{UltraNest}. The model weights are calculated using \texttt{arviz.compare} by stacking methods \citep{Yao2018Stacking}.}
\end{deluxetable*}

For reference, Table~\ref{tab:example-params} reports the fitted parameter summaries and uncertainty ranges for the absorbed power-law model used in this comparison.

\begin{deluxetable}{lcc}
\tablecaption{Parameter summaries for the absorbed power-law fit used in the HXMT Crab examples.\label{tab:example-params}}
\tablehead{
\colhead{Parameter} &
\colhead{MLE ($\pm 1\sigma$)} &
\colhead{Bayesian median ($\pm 1\sigma$)}
}
\startdata
$n_\mathrm{H}$ [$10^{22}$ cm$^{-2}$] & $0.3863 \pm 0.0223$ & $0.3873^{+0.0228}_{-0.0214}$ \\
$\alpha$                             & $2.1202 \pm 0.0026$ & $2.1204^{+0.0025}_{-0.0028}$ \\
$K$ [ph\,cm$^{-2}$\,s$^{-1}$\,keV$^{-1}$] & $8.952 \pm 0.060$ & $8.956^{+0.058}_{-0.062}$ \\
\enddata
\tablecomments{The Bayesian summaries correspond to the illustrative NUTS configuration used for the HXMT Crab examples. In the cutoff-power-law comparison, the additional cutoff energy is weakly constrained and can drift to values far above the fitted energy range.}
\end{deluxetable}

For these data, predictive criteria and MLE information criteria show no
decisive preference once uncertainty is accounted for (differences are small
compared to the reported standard errors). In the cutoff model, the cutoff
energy can drift to values far above the observed bandpass, effectively
reducing the model to an ordinary power law; in such settings, the simpler
model is typically preferred unless external information motivates the
additional parameter.

Further examples of \ELISA\ applications to more complex GRB and magnetar spectral analyses can be found in recent studies \citep[e.g.,][]{2025SCPMA..6871011Z,2026ApJ...999..159X,2026ApJ..1003..136X,2025ApJ...987L..38Z}.

\section{Conclusion}\label{sec:conclusion}

We introduced \ELISA, an open-source Python toolkit for forward-folded X/$\gamma$-ray
spectral analysis designed around a workflow-first Bayesian philosophy.
\ELISA\ unifies OGIP data handling, composable photon-space model construction, and a set
of likelihoods tailored to binned counting spectra, including widely used profile-background
treatments (W-stat and PG-stat).
Built on a \texttt{JAX} computational core, \ELISA\ provides just-in-time compiled model
evaluation and automatic differentiation, enabling both fast prototyping via
maximum-likelihood fits and practical posterior inference with modern samplers.
Beyond fitting, \ELISA\ treats model checking and comparison as first-class stages of the
analysis loop by integrating distributional calibration diagnostics based on PSIS-LOO
(PIT, Q--Q, and quantile residuals), posterior predictive checks, and predictive
information criteria (PSIS-LOO and WAIC), with evidence-based options available when
nested sampling is used.

The \textit{Insight}-HXMT Crab case study demonstrates how these components combine into a
reproducible, in the operational sense defined in Section~\ref{sec:overview:goals}, end-to-end analysis, from response-aware grouping and joint multi-instrument
likelihood construction to convergence assessment and calibrated diagnostic plots.
In particular, the example highlights the practical advantage of differentiable
forward-folding for gradient-informed sampling, translating into substantial gains in
effective sampling efficiency relative to commonly used ensemble workflows in this setting.

Several limitations and extensions motivate ongoing development.
As in other spectral-fitting ecosystems, inference and diagnostic interpretation remain
nontrivial in non-regular regimes (e.g., low counts, boundary constraints, and weakly
identified parameters), where simulation-based calibration and posterior predictive
diagnostics should be preferred over heuristic reduced-statistic summaries.
Interoperability with external non-\texttt{JAX} components (e.g., \XSPEC\ models) can
introduce additional overhead and relies on finite-difference sensitivity approximations,
which may reduce the benefits of gradient-based methods for some models.
Looking ahead, we plan to expand the curated set of reproducible examples and testing
assets, improve performance across a broader range of response geometries and hardware
backends, and further develop workflow support for rigorous sensitivity analyses and
model-averaging strategies when model uncertainty is non-negligible.
Together, these directions aim to make statistically principled Bayesian spectral
analysis increasingly routine, transparent, and reproducible for high-energy
astronomical applications.

\section*{Software Availability}\label{sec:availability}

The source code of \ELISA\ is hosted on GitHub at \url{https://github.com/wcxve/elisa} and is distributed on PyPI under the package name \texttt{astro-elisa} (install via \texttt{pip install astro-elisa}).
A frozen software release is also archived on Zenodo and should be cited as the versioned software record for this work (\citealt{ELISA}; DOI: \url{https://doi.org/10.5281/zenodo.11284455}).
The documentation and tutorials are hosted online via Read the Docs at \url{https://astro-elisa.rtfd.io}. \ELISA\ is licensed under GPL v3.0+, consistent with the public repository and archived release.

\begin{acknowledgments}

This work is supported by
the National Key R\&D Program of China (2021YFA0718500),
the Strategic Priority Research Program of the Chinese Academy of Sciences (Grant No. XDB0550300),
the National Natural Science Foundation of China (Grant No. 12273042), and
China's Space Origins Exploration Program.
This work made use of data from the \textit{Insight}-HXMT mission, funded by the CNSA and CAS. We appreciate the support from the GECAM team and the Insight-HXMT team. GECAM (Huairou-1) mission is supported by the Strategic Priority Research Program on Space Science (Grant No. XDA15360000) of Chinese Academy of Sciences.

\end{acknowledgments}

\appendix

\section{Profile background solutions for W-stat and PG-stat}\label{app:profile-background}

This appendix derives the per-channel profile background solutions used by the
W-stat and PG-stat likelihoods in Section~\ref{sec:stats:wstat} and
Section~\ref{sec:stats:pgstat}.
Because the profile optimization is separable across channels, we suppress the
dataset/channel indices $(d,j)$ and write $n\equiv N_{dj}$,
$m\equiv N^{\mathrm{off}}_{dj}$, $s\equiv s_{dj}(\Theta)$, $\alpha\equiv\alpha_d$,
and $b\equiv b_{dj}$.
In both cases, we enforce the physical constraint $b\ge 0$.

\subsection{W-stat (Poisson on/off)}\label{app:profile-background:wstat}

For W-stat, the per-channel generative model is
\begin{equation}
  n \sim \mathrm{Poisson}\!\left(s+\alpha b\right),
  \qquad
  m \sim \mathrm{Poisson}\!\left(b\right),
\end{equation}
where $b$ denotes the expected off-background counts and $\alpha b$ is the
corresponding background expectation in the on spectrum.
Up to constants that depend only on the observed data, the log likelihood is
\begin{equation}
  \ell(b) = n\ln\!\left(s+\alpha b\right) - (s+\alpha b) + m\ln b - b .
\end{equation}
Differentiating and setting $\partial \ell/\partial b=0$ yields
\begin{equation}
  \frac{\alpha n}{s+\alpha b} + \frac{m}{b} = \alpha + 1,
\end{equation}
which leads to a quadratic equation in $b$,
\begin{equation}
  \alpha(\alpha+1)b^2 + \left[(\alpha+1)s - \alpha(n+m)\right]b - ms = 0.
\end{equation}
The maximizer is obtained by the non-negative root
\begin{equation}
\begin{aligned}
  \hat{b}(s) &=
  \frac{c + \sqrt{c^2 + 4\alpha(\alpha+1)ms}}{2\alpha(\alpha+1)}, \\
  c &\equiv \alpha(n+m) - (\alpha+1)s,
\end{aligned}
\end{equation}
followed by truncation at $b\ge 0$ when necessary.
This expression reduces to $\hat{b}=m/(\alpha+1)$ when $n=0$, and to
$\hat{b}=\max\!\left(0,\,n/(\alpha+1)-s/\alpha\right)$ when $m=0$.

\subsection{PG-stat (Poisson on with Gaussian off)}\label{app:profile-background:pgstat}

For PG-stat, the on spectrum is Poisson while the background is supplied as a
Gaussian estimate with uncertainty:
\begin{equation}
  n \sim \mathrm{Poisson}\!\left(s+\alpha b\right),
  \qquad
  m \sim \mathcal{N}\!\left(b,\,\sigma^2\right),
\end{equation}
where $m$ and $\sigma$ correspond to the background estimate and its
per-channel standard deviation.
Up to data-dependent constants, the log likelihood is
\begin{equation}
  \ell(b) =
  n\ln\!\left(s+\alpha b\right) - (s+\alpha b) - \frac{(b-m)^2}{2\sigma^2}.
\end{equation}
Setting $\partial \ell/\partial b=0$ gives
\begin{equation}
  \frac{\alpha n}{s+\alpha b} - \alpha - \frac{b-m}{\sigma^2} = 0,
\end{equation}
which yields the quadratic equation
\begin{equation}
  \alpha b^2 + \left(s + \alpha^2\sigma^2 - \alpha m\right)b
  - \left(sm + \alpha\sigma^2(n-s)\right) = 0.
\end{equation}
Defining $v\equiv\sigma^2$, the non-negative profile solution can be written as
\begin{equation}
\begin{aligned}
  \hat{b}(s) &=
  \frac{c + \sqrt{c^2 + 4\alpha f}}{2\alpha},\\
  c &\equiv \alpha(m-\alpha v) - s,\\
  f &\equiv \alpha v n + (m-\alpha v)s,\\
\end{aligned}
\end{equation}
followed by truncation at $b\ge 0$.
For $n=0$, this reduces to $\hat{b}=\max(0,\,m-\alpha v)$.

\bibliographystyle{aasjournalv7}
\bibliography{refs}

@misc{ELISA,
  author    = {Xue, Wang-Chen and
               Xie, Sheng-Lun and
               Zheng, Chao and
               Xiong, Shao-Lin and
               Li, Xiao-Bo and
               Chen, Yong and
               Zhang, Peng and
               Zhang, Yan-Qiu and
               Liu, Jia-Cong and
               Wang, Chen-Wei and
               Tan, Wen-Jun and
               Wang, Yue and
               Zhang, Jin-Peng and
               Yu, Zheng-Hang},
  title     = {ELISA: Efficient Library for Spectral Analysis in
               High-Energy Astrophysics},
  month     = sep,
  year      = 2025,
  publisher = {Zenodo},
  doi       = {10.5281/zenodo.11284455},
  url       = {https://doi.org/10.5281/zenodo.11284455}
}

@INPROCEEDINGS{2000ASPC..216..591H,
       author = {{Houck}, J.~C. and {Denicola}, L.~A.},
        title = "{ISIS: An Interactive Spectral Interpretation System for High Resolution X-Ray Spectroscopy}",
    booktitle = {Astronomical Data Analysis Software and Systems IX},
         year = 2000,
       editor = {{Manset}, Nadine and {Veillet}, Christian and {Crabtree}, Dennis},
       series = {Astronomical Society of the Pacific Conference Series},
       volume = {216},
        month = jan,
        pages = {591},
       adsurl = {https://ui.adsabs.harvard.edu/abs/2000ASPC..216..591H}
}

@ARTICLE{2023A&A...678A.157D,
       author = {{Donath}, Axel and {Terrier}, R{\'e}gis and {Remy}, Quentin and {Sinha}, Atreyee and {Nigro}, Cosimo and {Pintore}, Fabio and {Kh{\'e}lifi}, Bruno and {Olivera-Nieto}, Laura and {Ruiz}, Jose Enrique and {Br{\"u}gge}, Kai and {Linhoff}, Maximilian and {Contreras}, Jose Luis and {Acero}, Fabio and {Aguasca-Cabot}, Arnau and {Berge}, David and {Bhattacharjee}, Pooja and {Buchner}, Johannes and {Boisson}, Catherine and {Carreto Fidalgo}, David and {Chen}, Andrew and {de Bony de Lavergne}, Mathieu and {de Miranda Cardoso}, Jos{\'e} Vinicius and {Deil}, Christoph and {F{\"u}{\ss}ling}, Matthias and {Funk}, Stefan and {Giunti}, Luca and {Hinton}, Jim and {Jouvin}, L{\'e}a and {King}, Johannes and {Lefaucheur}, Julien and {Lemoine-Goumard}, Marianne and {Lenain}, Jean-Philippe and {L{\'o}pez-Coto}, Rub{\'e}n and {Mohrmann}, Lars and {Morcuende}, Daniel and {Panny}, Sebastian and {Regeard}, Maxime and {Saha}, Lab and {Siejkowski}, Hubert and {Siemiginowska}, Aneta and {Sip{\H{o}}cz}, Brigitta M. and {Unbehaun}, Tim and {van Eldik}, Christopher and {Vuillaume}, Thomas and {Zanin}, Roberta},
        title = "{Gammapy: A Python package for gamma-ray astronomy}",
      journal = {\aap},
         year = 2023,
        month = oct,
       volume = {678},
          eid = {A157},
        pages = {A157},
          doi = {10.1051/0004-6361/202346488},
archivePrefix = {arXiv},
       eprint = {2308.13584},
 primaryClass = {astro-ph.IM},
       adsurl = {https://ui.adsabs.harvard.edu/abs/2023A&A...678A.157D}
}

@INPROCEEDINGS{2011ASPC..442..439S,
       author = {{Siemiginowska}, A. and {Kashyap}, V. and {Refsdal}, B. and {van Dyk}, D. and {Connors}, A. and {Park}, T.},
        title = "{pyblocxs: Bayesian Low-Counts X-ray Spectral Analysis in Sherpa}",
    booktitle = {Astronomical Data Analysis Software and Systems XX},
         year = 2011,
       editor = {{Evans}, I.~N. and {Accomazzi}, A. and {Mink}, D.~J. and {Rots}, A.~H.},
       series = {Astronomical Society of the Pacific Conference Series},
       volume = {442},
        month = jul,
        pages = {439},
       adsurl = {https://ui.adsabs.harvard.edu/abs/2011ASPC..442..439S}
}

@ARTICLE{2001ApJ...548..224V,
       author = {{van Dyk}, David A. and {Connors}, Alanna and {Kashyap}, Vinay L. and {Siemiginowska}, Aneta},
        title = "{Analysis of Energy Spectra with Low Photon Counts via Bayesian Posterior Simulation}",
      journal = {\apj},
         year = 2001,
        month = feb,
       volume = {548},
       number = {1},
        pages = {224-243},
          doi = {10.1086/318656},
archivePrefix = {arXiv},
       eprint = {astro-ph/0008170},
 primaryClass = {astro-ph},
       adsurl = {https://ui.adsabs.harvard.edu/abs/2001ApJ...548..224V}
}

@ARTICLE{2025SCPMA..6871011Z,
       author = {{Zheng}, Chao and {Xiong}, Shao-Lin and {Li}, Cheng-Kui and {Li}, Xiao-Bo and {Li}, Xu-Fang and {Li}, Gang and {Zhang}, Peng and {Xue}, Wang-Chen and {Zhang}, Yan-Qiu and {Liu}, Cong-Zhan and {Zhang}, Juan and {Li}, Bing and {Ge}, Ming-Yu and {Song}, Li-Ming and {Tan}, Wen-Jun and {Wang}, Chen-Wei and {Wang}, Yue and {Wang}, Jin-Peng and {Liu}, Jia-Cong and {Xie}, Sheng-Lun and {Cai}, Ce and {Xiao}, Shuo and {Zhang}, Wen-Long and {Guo}, Hao-Xuan and {Xu}, Yu-Peng and {Zhang}, Shu and {Zhang}, Shuang-Nan},
        title = "{Improved performance of Insight-HXMT/HE CsI detectors}",
      journal = {Science China Physics, Mechanics, and Astronomy},
         year = 2025,
        month = jul,
       volume = {68},
       number = {7},
          eid = {271011},
        pages = {271011},
          doi = {10.1007/s11433-025-2676-5},
       adsurl = {https://ui.adsabs.harvard.edu/abs/2025SCPMA..6871011Z}
}

@article{Kaastra2016OptBin,
  author        = {{Kaastra}, J.~S. and {Bleeker}, J.~A.~M.},
  title         = {{Optimal binning of X-ray spectra and response matrix design}},
  journal       = {\aap},
  year          = 2016,
  month         = mar,
  volume        = {587},
  eid           = {A151},
  pages         = {A151},
  doi           = {10.1051/0004-6361/201527395},
  archiveprefix = {arXiv},
  eprint        = {1601.05309},
  primaryclass  = {astro-ph.IM},
  adsurl        = {https://ui.adsabs.harvard.edu/abs/2016A&A...587A.151K}
}

@article{Wachter1979Wstat,
  author   = {{Wachter}, K. and {Leach}, R. and {Kellogg}, E.},
  title    = {{Parameter estimation in X-ray astronomy using maximum likelihood.}},
  journal  = {\apj},
  year     = 1979,
  month    = may,
  volume   = {230},
  pages    = {274-287},
  doi      = {10.1086/157084},
  adsurl   = {https://ui.adsabs.harvard.edu/abs/1979ApJ...230..274W}
}

@article{Cash1979Cstat,
  author   = {{Cash}, W.},
  title    = {{Parameter estimation in astronomy through application of the likelihood ratio.}},
  journal  = {\apj},
  year     = 1979,
  month    = mar,
  volume   = {228},
  pages    = {939-947},
  doi      = {10.1086/156922},
  adsurl   = {https://ui.adsabs.harvard.edu/abs/1979ApJ...228..939C}
}

@article{Wilks1938,
  author  = {Wilks, Samuel S.},
  title   = {The Large-Sample Distribution of the Likelihood Ratio for Testing Composite Hypotheses},
  journal = {The Annals of Mathematical Statistics},
  year    = {1938},
  volume  = {9},
  number  = {1},
  pages   = {60--62},
  doi     = {10.1214/aoms/1177732360},
  url     = {https://projecteuclid.org/euclid.aoms/1177732360}
}

@book{McCullagh1989GLM,
  author    = {McCullagh, Peter and Nelder, John A.},
  title     = {Generalized Linear Models},
  publisher = {Chapman \& Hall},
  address   = {London},
  edition   = {2},
  year      = {1989}
}

@article{Hoffman2014NUTS,
  author  = {Matthew D. Hoffman and Andrew Gelman},
  title   = {The No-U-Turn Sampler: Adaptively Setting Path Lengths in Hamiltonian Monte Carlo},
  journal = {Journal of Machine Learning Research},
  year    = {2014},
  volume  = {15},
  number  = {47},
  pages   = {1593--1623},
  url     = {http://jmlr.org/papers/v15/hoffman14a.html}
}

@article{Watanabe2010WAIC,
  author  = {Sumio Watanabe},
  title   = {Asymptotic Equivalence of Bayes Cross Validation and Widely Applicable Information Criterion in Singular Learning Theory},
  journal = {Journal of Machine Learning Research},
  year    = {2010},
  volume  = {11},
  number  = {116},
  pages   = {3571--3594},
  url     = {http://jmlr.org/papers/v11/watanabe10a.html}
}

@article{Vehtari2015LOO,
  author        = {{Vehtari}, Aki and {Gelman}, Andrew and {Gabry}, Jonah},
  title         = {{Practical Bayesian model evaluation using leave-one-out cross-validation and WAIC}},
  journal       = {arXiv e-prints},
  year          = 2015,
  month         = jul,
  eid           = {arXiv:1507.04544},
  pages         = {arXiv:1507.04544},
  doi           = {10.48550/arXiv.1507.04544},
  archiveprefix = {arXiv},
  eprint        = {1507.04544},
  primaryclass  = {stat.CO},
  adsurl        = {https://ui.adsabs.harvard.edu/abs/2015arXiv150704544V}
}

@article{Gelman1996PPC,
  author  = {Gelman, Andrew and Meng, Xiao-Li and Stern, Hal},
  title   = {Posterior predictive assessment of model fitness via realized discrepancies},
  journal = {Statistica Sinica},
  volume  = {6},
  pages   = {733--760},
  year    = {1996}
}

@article{Kumar2019Arviz,
  author  = {Kumar, Ravin and Carroll, Colin and Hartikainen, Ari and Martin, Osvaldo},
  title   = {ArviZ: a unified library for exploratory analysis of Bayesian models in Python},
  journal = {Journal of Open Source Software},
  volume  = {4},
  number  = {33},
  pages   = {1143},
  year    = {2019},
  doi     = {10.21105/joss.01143}
}

@inproceedings{Arnaud1996,
  author    = {Arnaud, Keith A.},
  title     = {{XSPEC}: The First Ten Years},
  booktitle = {Astronomical Data Analysis Software and Systems V},
  editor    = {Jacoby, G. H. and Barnes, J.},
  series    = {ASP Conf. Ser.},
  volume    = {101},
  pages     = {17},
  year      = {1996}
}

@inproceedings{Freeman2001Sherpa,
  author        = {{Freeman}, Peter and {Doe}, Stephen and {Siemiginowska}, Aneta},
  title         = {{Sherpa: a mission-independent data analysis application}},
  booktitle     = {Astronomical Data Analysis},
  year          = 2001,
  editor        = {{Starck}, Jean-Luc and {Murtagh}, Fionn D.},
  series        = {Society of Photo-Optical Instrumentation Engineers (SPIE) Conference Series},
  volume        = {4477},
  month         = nov,
  pages         = {76-87},
  doi           = {10.1117/12.447161},
  archiveprefix = {arXiv},
  eprint        = {astro-ph/0108426},
  primaryclass  = {astro-ph},
  adsurl        = {https://ui.adsabs.harvard.edu/abs/2001SPIE.4477...76F}
}

@inproceedings{Doe2007Sherpa,
  author    = {{Doe}, S. and {Nguyen}, D. and {Stawarz}, C. and {Refsdal}, B. and {Siemiginowska}, A. and {Burke}, D. and {Evans}, I. and {Evans}, J. and {McDowell}, J. and {Houck}, J. and {Nowak}, M.},
  title     = {{Developing Sherpa with Python}},
  booktitle = {Astronomical Data Analysis Software and Systems XVI},
  year      = 2007,
  editor    = {{Shaw}, R.~A. and {Hill}, F. and {Bell}, D.~J.},
  series    = {Astronomical Society of the Pacific Conference Series},
  volume    = {376},
  month     = oct,
  pages     = {543},
  adsurl    = {https://ui.adsabs.harvard.edu/abs/2007ASPC..376..543D}
}

@article{Siemiginowska2024Sherpa,
  author        = {{Siemiginowska}, Aneta and {Burke}, Douglas and {G{\"u}nther}, Hans Moritz and {Lee}, Nicholas P. and {McLaughlin}, Warren and {Principe}, David A. and {Cheer}, Harlan and {Fruscione}, Antonella and {Laurino}, Omar and {McDowell}, Jonathan and {Terrell}, Marie},
  title         = {{Sherpa: An Open-source Python Fitting Package}},
  journal       = {\apjs},
  year          = 2024,
  month         = oct,
  volume        = {274},
  number        = {2},
  eid           = {43},
  pages         = {43},
  doi           = {10.3847/1538-4365/ad7bab},
  archiveprefix = {arXiv},
  eprint        = {2409.10400},
  primaryclass  = {astro-ph.IM},
  adsurl        = {https://ui.adsabs.harvard.edu/abs/2024ApJS..274...43S}
}

@article{Vianello2015ThreeML,
  author        = {{Vianello}, Giacomo and {Lauer}, Robert J. and {Younk}, Patrick and {Tibaldo}, Luigi and {Burgess}, James M. and {Ayala}, Hugo and {Harding}, Patrick and {Hui}, Michelle and {Omodei}, Nicola and {Zhou}, Hao},
  title         = {{The Multi-Mission Maximum Likelihood framework (3ML)}},
  journal       = {arXiv e-prints},
  year          = 2015,
  month         = jul,
  eid           = {arXiv:1507.08343},
  pages         = {arXiv:1507.08343},
  doi           = {10.48550/arXiv.1507.08343},
  archiveprefix = {arXiv},
  eprint        = {1507.08343},
  primaryclass  = {astro-ph.HE},
  adsurl        = {https://ui.adsabs.harvard.edu/abs/2015arXiv150708343V}
}

@misc{Bradbury2018JAX,
  author = {James Bradbury and Roy Frostig and Peter Hawkins and Matthew James Johnson and Chris Leary and Dougal Maclaurin and George Necula and Adam Paszke and Jake Vander{P}las and Skye Wanderman-{M}ilne and Qiao Zhang},
  title  = {{JAX}: composable transformations of {P}ython+{N}um{P}y programs},
  url    = {http://github.com/jax-ml/jax},
  year   = {2018}
}

@article{Dupourque2024jaxspec,
  author        = {{Dupourqu{\'e}}, S. and {Barret}, D. and {Diez}, C.~M. and {Guillot}, S. and {Quintin}, E.},
  title         = {{jaxspec: A fast and robust Python library for X-ray spectral fitting}},
  journal       = {\aap},
  year          = 2024,
  month         = oct,
  volume        = {690},
  eid           = {A317},
  pages         = {A317},
  doi           = {10.1051/0004-6361/202451736},
  archiveprefix = {arXiv},
  eprint        = {2409.05757},
  primaryclass  = {astro-ph.IM},
  adsurl        = {https://ui.adsabs.harvard.edu/abs/2024A&A...690A.317D}
}

@article{Buchner2014BXA,
  author        = {{Buchner}, J. and {Georgakakis}, A. and {Nandra}, K. and {Hsu}, L. and {Rangel}, C. and {Brightman}, M. and {Merloni}, A. and {Salvato}, M. and {Donley}, J. and {Kocevski}, D.},
  title         = {{X-ray spectral modelling of the AGN obscuring region in the CDFS: Bayesian model selection and catalogue}},
  journal       = {\aap},
  year          = 2014,
  month         = apr,
  volume        = {564},
  eid           = {A125},
  pages         = {A125},
  doi           = {10.1051/0004-6361/201322971},
  archiveprefix = {arXiv},
  eprint        = {1402.0004},
  primaryclass  = {astro-ph.HE},
  adsurl        = {https://ui.adsabs.harvard.edu/abs/2014A&A...564A.125B}
}

@article{Buchner2021BXA,
  author   = {{Buchner}, Johannes},
  title    = {{Bayesian X-ray Analysis (BXA) v4.0}},
  journal  = {The Journal of Open Source Software},
  year     = 2021,
  month    = may,
  volume   = {6},
  number   = {61},
  eid      = {3045},
  pages    = {3045},
  doi      = {10.21105/joss.03045},
  adsurl   = {https://ui.adsabs.harvard.edu/abs/2021JOSS....6.3045B}
}

@article{Buchner2016UltraNest,
  author        = {{Buchner}, Johannes},
  title         = {{A statistical test for Nested Sampling algorithms}},
  journal       = {Statistics and Computing},
  year          = 2016,
  month         = jan,
  volume        = {26},
  number        = {1-2},
  pages         = {383-392},
  doi           = {10.1007/s11222-014-9512-y},
  archiveprefix = {arXiv},
  eprint        = {1407.5459},
  primaryclass  = {stat.CO},
  adsurl        = {https://ui.adsabs.harvard.edu/abs/2016S&C....26..383B}
}

@article{Buchner2019UltraNest,
  author        = {{Buchner}, Johannes},
  title         = {{Collaborative Nested Sampling: Big Data versus Complex Physical Models}},
  journal       = {\pasp},
  year          = 2019,
  month         = oct,
  volume        = {131},
  number        = {1004},
  pages         = {108005},
  doi           = {10.1088/1538-3873/aae7fc},
  archiveprefix = {arXiv},
  eprint        = {1707.04476},
  primaryclass  = {stat.CO},
  adsurl        = {https://ui.adsabs.harvard.edu/abs/2019PASP..131j8005B}
}

@article{Buchner2021UltraNest,
  author        = {{Buchner}, Johannes},
  title         = {{UltraNest - a robust, general purpose Bayesian inference engine}},
  journal       = {The Journal of Open Source Software},
  year          = 2021,
  month         = apr,
  volume        = {6},
  number        = {60},
  eid           = {3001},
  pages         = {3001},
  doi           = {10.21105/joss.03001},
  archiveprefix = {arXiv},
  eprint        = {2101.09604},
  primaryclass  = {stat.CO},
  adsurl        = {https://ui.adsabs.harvard.edu/abs/2021JOSS....6.3001B}
}

@misc{Kaastra2024SPEX,
  author    = {{Kaastra}, J.~S. and {Raassen}, A.~J.~J. and {de Plaa}, J. and {Gu}, Liyi},
  title     = {{SPEX X-ray spectral fitting package}},
  year      = 2024,
  month     = mar,
  eid       = {10.5281/zenodo.10822753},
  doi       = {10.5281/zenodo.10822753},
  version   = {3.08.00},
  publisher = {Zenodo},
  adsurl    = {https://ui.adsabs.harvard.edu/abs/2024zndo..10822753K}
}

@article{Gelman2020Workflow,
  author        = {{Gelman}, Andrew and {Vehtari}, Aki and {Simpson}, Daniel and {Margossian}, Charles C. and {Carpenter}, Bob and {Yao}, Yuling and {Kennedy}, Lauren and {Gabry}, Jonah and {B{\"u}rkner}, Paul-Christian and {Modr{\'a}k}, Martin},
  title         = {{Bayesian Workflow}},
  journal       = {arXiv e-prints},
  year          = 2020,
  month         = nov,
  eid           = {arXiv:2011.01808},
  pages         = {arXiv:2011.01808},
  doi           = {10.48550/arXiv.2011.01808},
  archiveprefix = {arXiv},
  eprint        = {2011.01808},
  primaryclass  = {stat.ME},
  adsurl        = {https://ui.adsabs.harvard.edu/abs/2020arXiv201101808G}
}

@article{protassov_statistics_2002,
  title      = {Statistics, {Handle} with {Care}: {Detecting} {Multiple} {Model} {Components} with the {Likelihood} {Ratio} {Test}},
  volume     = {571},
  issn       = {0004-637X},
  shorttitle = {Statistics, {Handle} with {Care}},
  url        = {https://ui.adsabs.harvard.edu/abs/2002ApJ...571..545P},
  doi        = {10.1086/339856},
  urldate    = {2025-08-12},
  journal    = {The Astrophysical Journal},
  author     = {Protassov, Rostislav and van Dyk, David A. and Connors, Alanna and Kashyap, Vinay L. and Siemiginowska, Aneta},
  month      = may,
  year       = {2002},
  note       = {Publisher: IOP
                ADS Bibcode: 2002ApJ...571..545P},
  pages      = {545--559}
}

@article{GECAM1,
  author  = {{Li}, X. Q. and Wen, X. Y. and An, Z. H. and Cai C. and Chang, Z. and Chen, G.},
  title   = {The Technology for Detection of Gamma-Ray Burst with GECAM Satellite},
  journal = {Radiation Detection Technology and Methods},
  doi     = {10.1007/s41605-021-00288-z},
  year    = {2022}
}

@article{GECAM2,
  author  = {An, Z. H. and Sun, X. L. and Zhang, D. L. and Yang, S. and Li, X. Q. and Wen, X. Y. and Gong, K. and Liang, X. H. and Liu, X. J. and Liu, Y. Q. and Li, Y. G. and Xiong, S. L. and Xu, Y. B. and Zhang, Fan and Zhao, X. Y. and Cai, C. and Chang, Z. and Chen, G. and Chen, C. and Du, Y. Y. and Feng, P. Y. and Gao, M. and Gao, R. and Guo, D. Y. and He, J. J. and Hou, D. J. and Li, C. Y. and Li, G. and Li, L. and Li, X. F. and Li, M. S. and Lu, F. J. and Lu, H. and Meng, B. and Peng, W. X. and Shi, F. and Wang, H. and Wang, J. Z. and Wang, Y. S. and Wang, H. Z. and Wen, X. and Xiao, S. and Xu, Y. P. and Yang, J. W. and Yi, Q. B. and Zhang, S. N. and Zhang, C. Y. and Zhang, C. M. and Zhang, Fei and Zhao, Y. and Zhou, X.},
  title   = {The design and performance of GRD onboard the GECAM satellite},
  journal = {Radiation Detection Technology and Methods},
  volume  = {6},
  number  = {1},
  pages   = {43-52},
  issn    = {2509-9949},
  doi     = {10.1007/s41605-021-00289-y},
  url     = {https://doi.org/10.1007/s41605-021-00289-y},
  year    = {2022},
  type    = {Journal Article}
}

@article{Zhang20HXMTOverview,
  title    = {Overview to the {Hard} {X}-ray {Modulation} {Telescope} ({Insight}-{HXMT}) {Satellite}},
  volume   = {63},
  issn     = {1674-7348, 1869-1927},
  url      = {http://link.springer.com/10.1007/s11433-019-1432-6},
  doi      = {10.1007/s11433-019-1432-6},
  language = {en},
  number   = {4},
  urldate  = {2022-06-09},
  journal  = {Science China Physics, Mechanics \& Astronomy},
  author   = {Zhang, Shuang-Nan and Li, TiPei and Lu, FangJun and Song, LiMing and Xu, YuPeng and Liu, CongZhan and Chen, Yong and Cao, XueLei and Bu, QingCui and Chang, Zhi and Chen, Gang and Chen, Li and Chen, TianXiang and Chen, YiBao and Chen, YuPeng and Cui, Wei and Cui, WeiWei and Deng, JingKang and Dong, YongWei and Du, YuanYuan and Fu, MinXue and Gao, GuanHua and Gao, He and Gao, Min and Ge, MingYu and Gu, YuDong and Guan, Ju and Gungor, Can and Guo, ChengCheng and Han, DaWei and Hu, Wei and Huang, Yue and Huo, Jia and Jia, ShuMei and Jiang, LuHua and Jiang, WeiChun and Jin, Jing and Jin, YongJie and Li, Bing and Li, ChengKui and Li, Gang and Li, MaoShun and Li, Wei and Li, Xian and Li, XiaoBo and Li, XuFang and Li, YanGuo and Li, ZiJian and Li, ZhengWei and Liang, XiaoHua and Liao, JinYuan and Liu, GuoQing and Liu, HongWei and Liu, ShaoZhen and Liu, XiaoJing and Liu, Yuan and Liu, YiNong and Lu, Bo and Lu, XueFeng and Luo, Tao and Ma, Xiang and Meng, Bin and Nang, Yi and Nie, JianYin and Ou, Ge and Qu, JinLu and Sai, Na and Shang, RenCheng and Shen, GuoHong and Sun, Liang and Tan, Ying and Tao, Lian and Tuo, YouLi and Wang, Chen and Wang, ChunQin and Wang, GuoFeng and Wang, HuanYu and Wang, Juan and Wang, WenShuai and Wang, YuSa and Wen, XiangYang and Wu, BaiYang and Wu, BoBing and Wu, Mei and Xiao, GuangCheng and Xiong, ShaoLin and Yan, LinLi and Yang, JiaWei and Yang, Sheng and Yang, YanJi and Yi, QiBin and Yuan, Bin and Zhang, AiMei and Zhang, ChunLei and Zhang, ChengMo and Zhang, Fan and Zhang, HongMei and Zhang, Juan and Zhang, Qiang and Zhang, ShenYi and Zhang, Shu and Zhang, Tong and Zhang, WanChang and Zhang, Wei and Zhang, WenZhao and Zhang, Yi and Zhang, YiFei and Zhang, YongJie and Zhang, Yue and Zhang, Zhao and Zhang, Zhi and Zhang, ZiLiang and Zhao, HaiSheng and Zhao, XiaoFan and Zheng, ShiJie and Zhou, JianFeng and Zhu, YuXuan and Zhu, Yue and Zhuang, RenLin},
  month    = apr,
  year     = {2020},
  note     = {TLDR: An overview of the HXMT mission, including payload, core sciences, ground calibration/facility, ground segment, data archive, software, in-orbit performance, calibration, background model, observations and some preliminary results are given.},
  pages    = {249502}
}

@article{GBM_meegan_09,
  author        = {{Meegan}, Charles and {Lichti}, Giselher and {Bhat}, P.~N. and {Bissaldi}, Elisabetta and {Briggs}, Michael S. and {Connaughton}, Valerie and {Diehl}, Roland and {Fishman}, Gerald and {Greiner}, Jochen and {Hoover}, Andrew S. and {van der Horst}, Alexander J. and {von Kienlin}, Andreas and {Kippen}, R. Marc and {Kouveliotou}, Chryssa and {McBreen}, Sheila and {Paciesas}, W.~S. and {Preece}, Robert and {Steinle}, Helmut and {Wallace}, Mark S. and {Wilson}, Robert B. and {Wilson-Hodge}, Colleen},
  title         = {{The Fermi Gamma-ray Burst Monitor}},
  journal       = {\apj},
  year          = 2009,
  month         = sep,
  volume        = {702},
  number        = {1},
  pages         = {791-804},
  doi           = {10.1088/0004-637X/702/1/791},
  archiveprefix = {arXiv},
  eprint        = {0908.0450},
  primaryclass  = {astro-ph.IM},
  adsurl        = {https://ui.adsabs.harvard.edu/abs/2009ApJ...702..791M}
}

@article{SVM_GRM_Dong2010,
  author        = {{Dong}, Yongwei and {Wu}, Bobing and {Li}, Yanguo and {Zhang}, Yongjie and {Zhang}, Shuangnan},
  title         = {{SVOM gamma ray monitor}},
  journal       = {Science China Physics, Mechanics, and Astronomy},
  year          = 2010,
  month         = jan,
  volume        = {53},
  number        = {1},
  pages         = {40-42},
  doi           = {10.1007/s11433-010-0011-7},
  archiveprefix = {arXiv},
  eprint        = {0907.2768},
  primaryclass  = {astro-ph.IM},
  adsurl        = {https://ui.adsabs.harvard.edu/abs/2010SCPMA..53S..40D}
}

@article{SVOM_1,
  author        = {{Wei}, J. and {Cordier}, B. and {Antier}, S. and {Antilogus}, P. and {Atteia}, J. -L. and {Bajat}, A. and {Basa}, S. and {Beckmann}, V. and {Bernardini}, M.~G. and {Boissier}, S. and {Bouchet}, L. and {Burwitz}, V. and {Claret}, A. and {Dai}, Z. -G. and {Daigne}, F. and {Deng}, J. and {Dornic}, D. and {Feng}, H. and {Foglizzo}, T. and {Gao}, H. and {Gehrels}, N. and {Godet}, O. and {Goldwurm}, A. and {Gonzalez}, F. and {Gosset}, L. and {G{\"o}tz}, D. and {Gouiffes}, C. and {Grise}, F. and {Gros}, A. and {Guilet}, J. and {Han}, X. and {Huang}, M. and {Huang}, Y. -F. and {Jouret}, M. and {Klotz}, A. and {La Marle}, O. and {Lachaud}, C. and {Le Floch}, E. and {Lee}, W. and {Leroy}, N. and {Li}, L. -X. and {Li}, S.~C. and {Li}, Z. and {Liang}, E. -W. and {Lyu}, H. and {Mercier}, K. and {Migliori}, G. and {Mochkovitch}, R. and {O'Brien}, P. and {Osborne}, J. and {Paul}, J. and {Perinati}, E. and {Petitjean}, P. and {Piron}, F. and {Qiu}, Y. and {Rau}, A. and {Rodriguez}, J. and {Schanne}, S. and {Tanvir}, N. and {Vangioni}, E. and {Vergani}, S. and {Wang}, F. -Y. and {Wang}, J. and {Wang}, X. -G. and {Wang}, X. -Y. and {Watson}, A. and {Webb}, N. and {Wei}, J.~J. and {Willingale}, R. and {Wu}, C. and {Wu}, X. -F. and {Xin}, L. -P. and {Xu}, D. and {Yu}, S. and {Yu}, W. -F. and {Yu}, Y. -W. and {Zhang}, B. and {Zhang}, S. -N. and {Zhang}, Y. and {Zhou}, X.~L.},
  title         = {{The Deep and Transient Universe in the SVOM Era: New Challenges and Opportunities - Scientific prospects of the SVOM mission}},
  journal       = {arXiv e-prints},
  year          = 2016,
  month         = oct,
  eid           = {arXiv:1610.06892},
  pages         = {arXiv:1610.06892},
  doi           = {10.48550/arXiv.1610.06892},
  archiveprefix = {arXiv},
  eprint        = {1610.06892},
  primaryclass  = {astro-ph.IM},
  adsurl        = {https://ui.adsabs.harvard.edu/abs/2016arXiv161006892W}
}

@article{2025arXiv251215162L,
  author        = {{Li}, Rui-Zhi and {Mao}, Jirong and {Yang}, Yuan-Pei and {Wang}, Bo-Ting and {Song}, Fei-Fan and {Xin}, Yu-Xin and {Bai}, Jin-Ming},
  title         = {{Time Evolution of Optical Darkness in GRB Afterglow: The Case of GRB 240825A}},
  journal       = {arXiv e-prints},
  year          = 2025,
  month         = dec,
  eid           = {arXiv:2512.15162},
  pages         = {arXiv:2512.15162},
  doi           = {10.48550/arXiv.2512.15162},
  archiveprefix = {arXiv},
  eprint        = {2512.15162},
  primaryclass  = {astro-ph.HE},
  adsurl        = {https://ui.adsabs.harvard.edu/abs/2025arXiv251215162L}
}

@article{2025ApJ...993...24W,
  author        = {{Wang}, Yue and {Wang}, Chen-Wei and {Xiong}, Shao-Lin and {Xiao}, Shuo and {Zhang}, Yan-Qiu and {Xie}, Sheng-Lun and {Lin}, Lin and {Yang}, Yuan-Pei and {Guo}, Hao-Xuan and {Cai}, Ce and {Huang}, Yue and {Li}, Cheng-Kui and {Li}, Bing and {Li}, Xiao-Bo and {Liu}, Jia-Cong and {Ma}, Xiang and {Song}, Li-Ming and {Tan}, Wen-Jun and {Wang}, Ping and {Xue}, Wang-Chen and {Yi}, Shu-Xu and {Yu}, Yun-Wei and {Yu}, Zheng-Hang and {Zhang}, Jin-Peng and {Zhang}, Peng and {Zhang}, Shuang-Nan and {Zhang}, Wen-Long and {Zhang}, Zhen and {Zhao}, Xiao-Yun and {Zheng}, Chao and {Zheng}, Shi-Jie},
  title         = {{Unveil a Peculiar Light-curve Pattern of Magnetar Burst with GECAM Observations of SGR J1935+2154}},
  journal       = {\apj},
  year          = 2025,
  month         = nov,
  volume        = {993},
  number        = {1},
  eid           = {24},
  pages         = {24},
  doi           = {10.3847/1538-4357/ae0cbe},
  archiveprefix = {arXiv},
  eprint        = {2510.11385},
  primaryclass  = {astro-ph.HE},
  adsurl        = {https://ui.adsabs.harvard.edu/abs/2025ApJ...993...24W}
}

@ARTICLE{2026ApJ...999..159X,
       author = {{Xie}, Sheng-Lun and {Chen}, Aming and {Yu}, Yun-Wei and {Xiong}, Shao-Lin and {Feng}, Hua and {Zhang}, Shuang-Nan and {Dai}, Zi-Gao and {Xue}, Wang-Chen and {Ge}, Ming-Yu and {Li}, Xiao-Bo and {Liu}, Liang-Duan and {Liu}, Jia-Cong and {Tan}, Wen-Jun and {Wang}, Chen-Wei and {Yi}, Shu-Xu and {Zhang}, Peng and {Zhang}, Yan-Qiu and {Zhang}, Zhen and {Zheng}, Chao and {Zheng}, Xiao-Ping},
        title = "{Eclipsed X-Ray Bursts from Magnetar SGR J1935+2154 and the Fireball Measurements}",
      journal = {\apj},
         year = 2026,
        month = mar,
       volume = {999},
       number = {2},
          eid = {159},
        pages = {159},
          doi = {10.3847/1538-4357/ae433d},
archivePrefix = {arXiv},
       eprint = {2510.24075},
 primaryClass = {astro-ph.HE},
       adsurl = {https://ui.adsabs.harvard.edu/abs/2026ApJ...999..159X}
}

@ARTICLE{2026ApJ..1003..136X,
       author = {{Xie}, Sheng-Lun and {Yu}, Yun-Wei and {Xiong}, Shao-Lin and {Xue}, Wang-Chen},
        title = "{Spectral Evidence of Heavy Nuclei from the Neutron Star Crust in Magnetar Bursts}",
      journal = {\apj},
         year = 2026,
        month = jun,
       volume = {1003},
       number = {2},
          eid = {136},
        pages = {136},
          doi = {10.3847/1538-4357/ae6665},
archivePrefix = {arXiv},
       eprint = {2604.24750},
 primaryClass = {astro-ph.HE},
       adsurl = {https://ui.adsabs.harvard.edu/abs/2026ApJ..1003..136X}
}

@article{2025arXiv251015816W,
  author        = {{Wang}, Chen-Wei and {Zhang}, Peng and {Xiong}, Shao-Lin and {Huang}, Yue and {Tan}, Wen-Jun and {Yu}, Zheng-Hang and {Wang}, Yue and {Xue}, Wang-Chen and {Zheng}, Chao and {Guo}, Hao-Xuan and {Cai}, Ce and {Dong}, Yong-Wei and {He}, Jiang and {Li}, Cheng-Kui and {Li}, Xiao-Bo and {Liu}, Jia-Cong and {Luo}, Xing-Hao and {Ma}, Xiang and {Moradi}, Rahim and {Ren}, Yang-Zhao and {Song}, Li-Ming and {Wang}, Ping and {Wang}, Jin and {Wu}, Bo-Bing and {Xiao}, Shuo and {Xie}, Sheng-Lun and {Yi}, Shu-Xu and {Zao}, Xue-Yuan and {Zhao}, Xiao-Yun and {Zhang}, Li and {Zhang}, Shuang-Nan and {Zhang}, Yan-Qiu and {Zheng}, Shi-Jie},
  title         = {{BREAKFAST: A Framework for general joint BA duty and follow-up guidance of multiple $γ$-ray monitors}},
  journal       = {arXiv e-prints},
  year          = 2025,
  month         = oct,
  eid           = {arXiv:2510.15816},
  pages         = {arXiv:2510.15816},
  doi           = {10.48550/arXiv.2510.15816},
  archiveprefix = {arXiv},
  eprint        = {2510.15816},
  primaryclass  = {astro-ph.IM},
  adsurl        = {https://ui.adsabs.harvard.edu/abs/2025arXiv251015816W}
}

@article{2025arXiv250926283Z,
  author        = {{Zhang}, Jin-Peng and {Wang}, Chen-Wei and {Yu}, Zheng-Hang and {Xiong}, Shao-Lin and {Yi}, Shu-Xu and {Liu}, Jia-Cong and {Xue}, Wang-Chen and {Tan}, Wen-Jun and {Zhang}, Zi-Rui and {Moradi}, Rahim and {Guo}, Hao-Xuan and {Zheng}, Chao and {Zhang}, Yan-Qiu and {Wang}, Yue and {Xie}, Sheng-Lun and {Zhang}, Peng and {Ren}, Yang-Zhao and {Li}, Cheng-Kui and {Li}, Xiao-Bo and {Cai}, Ce and {Xiao}, Shuo and {Song}, Li-Ming and {Zhang}, Shuang-Nan},
  title         = {{On the Ultra-Long Gamma-Ray Transient GRB 250702B/EP250702}},
  journal       = {arXiv e-prints},
  year          = 2025,
  month         = sep,
  eid           = {arXiv:2509.26283},
  pages         = {arXiv:2509.26283},
  doi           = {10.48550/arXiv.2509.26283},
  archiveprefix = {arXiv},
  eprint        = {2509.26283},
  primaryclass  = {astro-ph.HE},
  adsurl        = {https://ui.adsabs.harvard.edu/abs/2025arXiv250926283Z}
}

@article{2025ApJ...987L...7L,
  author   = {{Liao}, Tong-Lei and {Xiao}, Shuo and {Xiong}, Shao-Lin and {Zhang}, Yan-Qiu and {Xue}, Wang-Chen and {Yang}, Jiao-Jiao and {Jiang}, Zheng-Huo and {Yi}, Shu-Xu and {Zhang}, Zhen and {Zhang}, Shuang-Nan},
  title    = {{Candidate Spectral Lines at 6 and 15 MeV from GRB 200829A and GRB 160905A}},
  journal  = {\apjl},
  year     = 2025,
  month    = jul,
  volume   = {987},
  number   = {1},
  eid      = {L7},
  pages    = {L7},
  doi      = {10.3847/2041-8213/ade4b8},
  adsurl   = {https://ui.adsabs.harvard.edu/abs/2025ApJ...987L...7L}
}

@article{2025ApJ...983..130L,
  author   = {{Liao}, Tong-Lei and {Zhang}, Yan-Qiu and {Xue}, Wang-Chen and {Xiao}, Shuo and {Xiong}, Shao-Lin and {Wang}, Xi-Lu and {Feng}, Hua and {Zheng}, Chao and {Wang}, Yue and {Liu}, Jia-Cong and {Wang}, Chen-Wei and {Yang}, Jiao-Jiao and {Hong}, Mei-Xin and {Jiang}, Zheng-Huo and {Lai}, Yang and {Luo}, Xi-Hong},
  title    = {{Spectral Line Search in Bright Gamma-Ray Bursts Observed by Fermi/GBM}},
  journal  = {\apj},
  year     = 2025,
  month    = apr,
  volume   = {983},
  number   = {2},
  eid      = {130},
  pages    = {130},
  doi      = {10.3847/1538-4357/adc00b},
  adsurl   = {https://ui.adsabs.harvard.edu/abs/2025ApJ...983..130L}
}

@article{2025ApJ...987L..38Z,
  author        = {{Zhang}, Yan-Qiu and {Xue}, Wang-Chen and {Zhang}, Jin-Peng and {Cai}, Ce and {Xiong}, Shao-Lin and {Li}, Cheng-Kui and {Liu}, Yuan and {Wang}, Chen-Wei and {Guo}, Hao-Xuan and {Xiao}, Shuo and {Tan}, Wen-Jun and {Zheng}, Chao and {Liu}, Jia-Cong and {Xie}, Sheng-Lun and {Zhang}, Peng and {Zhang}, Wen-Long and {Wang}, Yue and {Yu}, Zheng-Hang and {Ren}, Yang-Zhao and {Wang}, Ping and {Huang}, Yue and {Li}, Xiao-Bo and {Zhao}, Xiao-Yun and {Zheng}, Shi-Jie and {Zhang}, Zhen and {Yi}, Shu-Xu and {Song}, Li-Ming and {Zhang}, Shuang-Nan},
  title         = {{Multi-instrument Search for Gamma-Ray Counterpart of X-Ray Transients Detected by EP/WXT}},
  journal       = {\apjl},
  year          = 2025,
  month         = jul,
  volume        = {987},
  number        = {2},
  eid           = {L38},
  pages         = {L38},
  doi           = {10.3847/2041-8213/ade0aa},
  archiveprefix = {arXiv},
  eprint        = {2506.05920},
  primaryclass  = {astro-ph.HE},
  adsurl        = {https://ui.adsabs.harvard.edu/abs/2025ApJ...987L..38Z}
}

@incollection{Buchner23SpecStat,
  title     = {Statistical {{Aspects}} of {{X-ray Spectral Analysis}}},
  booktitle = {Handbook of {{X-ray}} and {{Gamma-ray Astrophysics}}},
  author    = {Buchner, Johannes and Boorman, Peter},
  year      = 2023,
  month     = jan,
  pages     = {150},
  doi       = {10.1007/978-981-16-4544-0_175-1}
}

@article{Humphrey2009Chi2Bias,
  author        = {{Humphrey}, Philip J. and {Liu}, Wenhao and {Buote}, David A.},
  title         = {{{\ensuremath{\chi}}$^{2}$ and Poissonian Data: Biases Even in the High-Count Regime and How to Avoid Them}},
  journal       = {\apj},
  year          = 2009,
  month         = mar,
  volume        = {693},
  number        = {1},
  pages         = {822-829},
  doi           = {10.1088/0004-637X/693/1/822},
  archiveprefix = {arXiv},
  eprint        = {0811.2796},
  primaryclass  = {astro-ph},
  adsurl        = {https://ui.adsabs.harvard.edu/abs/2009ApJ...693..822H}
}

@article{phan2019composable,
  title   = {Composable Effects for Flexible and Accelerated Probabilistic Programming in NumPyro},
  author  = {Phan, Du and Pradhan, Neeraj and Jankowiak, Martin},
  journal = {arXiv preprint arXiv:1912.11554},
  year    = {2019}
}

@article{bingham2019pyro,
  author  = {Eli Bingham and
             Jonathan P. Chen and
             Martin Jankowiak and
             Fritz Obermeyer and
             Neeraj Pradhan and
             Theofanis Karaletsos and
             Rohit Singh and
             Paul A. Szerlip and
             Paul Horsfall and
             Noah D. Goodman},
  title   = {Pyro: Deep Universal Probabilistic Programming},
  journal = {J. Mach. Learn. Res.},
  volume  = {20},
  pages   = {28:1--28:6},
  year    = {2019},
  url     = {http://jmlr.org/papers/v20/18-403.html}
}

@misc{cabezas2024blackjax,
  title         = {BlackJAX: Composable {B}ayesian inference in {JAX}},
  author        = {Alberto Cabezas and Adrien Corenflos and Junpeng Lao and Rémi Louf},
  year          = {2024},
  eprint        = {2402.10797},
  archiveprefix = {arXiv},
  primaryclass  = {cs.MS}
}

@article{Foreman-Mackey13emcee,
  title      = {Emcee : {{The MCMC Hammer}}},
  shorttitle = {Emcee},
  author     = {{Foreman-Mackey}, Daniel and Hogg, David W. and Lang, Dustin and Goodman, Jonathan},
  year       = 2013,
  month      = mar,
  journal    = {Publications of the Astronomical Society of the Pacific},
  volume     = {125},
  number     = {925},
  pages      = {306--312},
  issn       = {00046280, 15383873},
  doi        = {10.1086/670067}
}

@article{Karamanis2021zeus,
  author        = {{Karamanis}, Minas and {Beutler}, Florian and {Peacock}, John A.},
  title         = {{zeus: a PYTHON implementation of ensemble slice sampling for efficient Bayesian parameter inference}},
  journal       = {\mnras},
  year          = 2021,
  month         = dec,
  volume        = {508},
  number        = {3},
  pages         = {3589-3603},
  doi           = {10.1093/mnras/stab2867},
  archiveprefix = {arXiv},
  eprint        = {2105.03468},
  primaryclass  = {astro-ph.IM},
  adsurl        = {https://ui.adsabs.harvard.edu/abs/2021MNRAS.508.3589K}
}

@article{Albert2020JAXNS,
  author        = {{Albert}, Joshua G.},
  title         = {{JAXNS: a high-performance nested sampling package based on JAX}},
  journal       = {arXiv e-prints},
  year          = 2020,
  month         = dec,
  eid           = {arXiv:2012.15286},
  pages         = {arXiv:2012.15286},
  doi           = {10.48550/arXiv.2012.15286},
  archiveprefix = {arXiv},
  eprint        = {2012.15286},
  primaryclass  = {astro-ph.IM},
  adsurl        = {https://ui.adsabs.harvard.edu/abs/2020arXiv201215286A}
}

@article{Albert2023JAXNS,
  author        = {{Albert}, Joshua G.},
  title         = {{Phantom-Powered Nested Sampling}},
  journal       = {arXiv e-prints},
  year          = 2023,
  month         = dec,
  eid           = {arXiv:2312.11330},
  pages         = {arXiv:2312.11330},
  doi           = {10.48550/arXiv.2312.11330},
  archiveprefix = {arXiv},
  eprint        = {2312.11330},
  primaryclass  = {astro-ph.IM},
  adsurl        = {https://ui.adsabs.harvard.edu/abs/2023arXiv231211330A}
}

@article{Lange2023NAUTILUS,
  author        = {{Lange}, Johannes U.},
  title         = {{NAUTILUS: boosting Bayesian importance nested sampling with deep learning}},
  journal       = {\mnras},
  year          = 2023,
  month         = oct,
  volume        = {525},
  number        = {2},
  pages         = {3181-3194},
  doi           = {10.1093/mnras/stad2441},
  archiveprefix = {arXiv},
  eprint        = {2306.16923},
  primaryclass  = {astro-ph.IM},
  adsurl        = {https://ui.adsabs.harvard.edu/abs/2023MNRAS.525.3181L}
}

@article{Livingstone2022BarkerProposal,
  author   = {Livingstone, Samuel and Zanella, Giacomo},
  title    = {The Barker Proposal: Combining Robustness and Efficiency in Gradient-Based MCMC},
  journal  = {Journal of the Royal Statistical Society Series B: Statistical Methodology},
  volume   = {84},
  number   = {2},
  pages    = {496-523},
  year     = {2022},
  month    = {01},
  issn     = {1369-7412},
  doi      = {10.1111/rssb.12482},
  url      = {https://doi.org/10.1111/rssb.12482},
  eprint   = {https://academic.oup.com/jrsssb/article-pdf/84/2/496/49322274/jrsssb_84_2_496.pdf}
}

@inbook{Zhu2019SAMCMC,
  author    = {Zhu, Michael H.},
  title     = {Sample adaptive MCMC},
  year      = {2019},
  booktitle = {Proceedings of the 33rd International Conference on Neural Information Processing Systems},
  articleno = {813},
  numpages  = {12},
  doi       = {10.5555/3454287.3455100}
}

@article{Goodman2010AIES,
  author   = {{Goodman}, Jonathan and {Weare}, Jonathan},
  title    = {{Ensemble samplers with affine invariance}},
  journal  = {Communications in Applied Mathematics and Computational Science},
  year     = 2010,
  month    = jan,
  volume   = {5},
  number   = {1},
  pages    = {65-80},
  doi      = {10.2140/camcos.2010.5.65},
  adsurl   = {https://ui.adsabs.harvard.edu/abs/2010CAMCS...5...65G}
}

@article{Karamanis2020ESS,
  author        = {{Karamanis}, Minas and {Beutler}, Florian},
  title         = {{Ensemble Slice Sampling: Parallel, black-box and gradient-free inference for correlated \& multimodal distributions}},
  journal       = {arXiv e-prints},
  year          = 2020,
  month         = feb,
  eid           = {arXiv:2002.06212},
  pages         = {arXiv:2002.06212},
  doi           = {10.48550/arXiv.2002.06212},
  archiveprefix = {arXiv},
  eprint        = {2002.06212},
  primaryclass  = {stat.ML},
  adsurl        = {https://ui.adsabs.harvard.edu/abs/2020arXiv200206212K}
}

@incollection{Neal2011HMC,
  author    = {{Neal}, Radford},
  title     = {{MCMC Using Hamiltonian Dynamics}},
  booktitle = {Handbook of Markov Chain Monte Carlo},
  year      = 2011,
  pages     = {113-162},
  doi       = {10.1201/b10905},
  adsurl    = {https://ui.adsabs.harvard.edu/abs/2011hmcm.book..113N}
}

@article{Betancourt2013HMC,
  author        = {{Betancourt}, M.~J. and {Girolami}, Mark},
  title         = {{Hamiltonian Monte Carlo for Hierarchical Models}},
  journal       = {arXiv e-prints},
  year          = 2013,
  month         = dec,
  eid           = {arXiv:1312.0906},
  pages         = {arXiv:1312.0906},
  doi           = {10.48550/arXiv.1312.0906},
  archiveprefix = {arXiv},
  eprint        = {1312.0906},
  primaryclass  = {stat.ME},
  adsurl        = {https://ui.adsabs.harvard.edu/abs/2013arXiv1312.0906B}
}

@inproceedings{Skilling2004NS,
  author    = {{Skilling}, John},
  title     = {{Nested Sampling}},
  booktitle = {Bayesian Inference and Maximum Entropy Methods in Science and Engineering: 24th International Workshop on Bayesian Inference and Maximum Entropy Methods in Science and Engineering},
  year      = 2004,
  editor    = {{Fischer}, Rainer and {Preuss}, Roland and {Toussaint}, Udo Von},
  series    = {American Institute of Physics Conference Series},
  volume    = {735},
  month     = nov,
  publisher = {AIP},
  pages     = {395-405},
  doi       = {10.1063/1.1835238},
  adsurl    = {https://ui.adsabs.harvard.edu/abs/2004AIPC..735..395S}
}

@article{Vehtari2021Rhat,
  author        = {{Vehtari}, Aki and {Gelman}, Andrew and {Simpson}, Daniel and {Carpenter}, Bob and {B{\"u}rkner}, Paul-Christian},
  title         = {{Rank-normalization, folding, and localization: An improved R-hat for assessing convergence of MCMC (with Discussion)}},
  journal       = {Bayesian Analysis},
  year          = 2021,
  month         = jun,
  volume        = {16},
  number        = {2},
  pages         = {667-718},
  doi           = {10.1214/20-BA1221},
  archiveprefix = {arXiv},
  eprint        = {1903.08008},
  primaryclass  = {stat.CO},
  adsurl        = {https://ui.adsabs.harvard.edu/abs/2021BayAn..16..667V}
}

@article{Warton17PITtrap,
  title   = {The {{PIT-trap}}---{{A}} ``model-Free'' Bootstrap Procedure for Inference about Regression Models with Discrete, Multivariate Responses},
  author  = {Warton, David I. and Thibaut, Lo{\"i}c and Wang, Yi Alice},
  year    = 2017,
  month   = jul,
  journal = {PLoS ONE},
  volume  = {12},
  pages   = {e0181790},
  doi     = {10.1371/journal.pone.0181790}
}

@article{Dunn1996RQR,
  author    = {Peter K. Dunn and Gordon K. Smyth},
  title     = {Randomized Quantile Residuals},
  journal   = {Journal of Computational and Graphical Statistics},
  volume    = {5},
  number    = {3},
  pages     = {236--244},
  year      = {1996},
  publisher = {Taylor \& Francis},
  doi       = {10.1080/10618600.1996.10474708}
}

@article{Sailynoja22PITECDF,
  title   = {Graphical Test for Discrete Uniformity and Its Applications in Goodness-of-Fit Evaluation and Multiple Sample Comparison},
  author  = {S{\"a}ilynoja, Teemu and B{\"u}rkner, Paul-Christian and Vehtari, Aki},
  year    = 2022,
  month   = apr,
  journal = {Statistics and Computing},
  volume  = {32},
  number  = {2},
  pages   = {32},
  issn    = {0960-3174, 1573-1375},
  doi     = {10.1007/s11222-022-10090-6}
}

@article{Kass1995BayesFactors,
  title     = {Bayes Factors},
  author    = {Kass, Robert E. and Raftery, Adrian E.},
  journal   = {Journal of the American Statistical Association},
  volume    = {90},
  number    = {430},
  pages     = {773--795},
  year      = {1995},
  publisher = {Taylor \& Francis},
  doi       = {10.1080/01621459.1995.10476572}
}

@article{Hoeting1999BMA,
  title   = {Bayesian Model Averaging: A Tutorial},
  author  = {Hoeting, Jennifer A. and Madigan, David and Raftery, Adrian E. and Volinsky, Chris T.},
  journal = {Statistical Science},
  volume  = {14},
  number  = {4},
  pages   = {382--401},
  year    = {1999},
  doi     = {10.1214/ss/1009212519}
}

@article{Yao2018Stacking,
  title   = {Using Stacking to Average {Bayesian} Predictive Distributions},
  author  = {Yao, Yuling and Vehtari, Aki and Simpson, Daniel and Gelman, Andrew},
  journal = {Bayesian Analysis},
  volume  = {13},
  number  = {3},
  pages   = {917--1007},
  year    = {2018},
  doi     = {10.1214/17-BA1091}
}

\end{document}